\documentclass[12pt,a4paper]{article}
    \usepackage[format=hang]{caption}
    \usepackage{epsfig,amssymb,amsmath,graphicx,subcaption,verbatim,hyperref,ulem,bm,float,tensor,tikz}
    \usepackage{blkarray}
    \usepackage{xcolor}
    \usepackage{tikz}
    \usetikzlibrary{decorations.markings}
    \usepackage{amsthm}

    \definecolor{BrickRed}{RGB}{203, 65, 84}
    \definecolor{Violet}{HTML}{EE82EE}
    \definecolor{OliveGreen}{HTML}{556B2F}
    \definecolor{RoyalBlue}{RGB}{25,41,88}
    
    \hypersetup{colorlinks=true, linkcolor=BrickRed, citecolor=Violet, filecolor=OliveGreen, urlcolor=RoyalBlue, filebordercolor={.8 .8 1}, urlbordercolor={.8 .8 0}}

    \newcommand{\bZ}{\mathbb{Z}}
    
    \newcommand{\be}{\begin{equation}}
    \newcommand{\ee}{\end{equation}}

    \font\mybb=msbm10 at 11pt
    \def\bb#1{\hbox{\mybb#1}}
    \def\bZ {\bb{Z}}

    \DeclareMathOperator{\im}{Im}

\newcommand{\mC}{\mathcal{C}}
\newcommand{\mB}{\mathcal{B}}
\newcommand{\mP}{\mathcal{P}}
\newcommand{\mX}{\mathcal{X}}
\newcommand{\mZ}{\mathcal{Z}}

\newcommand{\Z}{\mathbb{Z}}
\newcommand{\R}{\mathbb{R}}
\newcommand{\one}{\mathbf{1}}
\def\mb#1{\mathbf{#1}}

\renewcommand{\det}{\text{det}}
    
 \def\wloop#1{
\scalebox{0.7}{
\begin{tikzpicture}[baseline=-0.65ex,scale=1.3]
 \draw (0,0) ellipse [x radius=10pt, y radius=4pt];
 \draw [->,thin] (0,-4pt)--(2pt,-4pt);
 \node at (4pt,-8pt) {$#1$};
\end{tikzpicture}
}}

\def\wloopVtc#1{
\scalebox{0.9}{
\begin{tikzpicture}[baseline=-0.65ex,scale=1]
 \draw (0,0) ellipse [x radius=4pt, y radius=10pt];
 \draw [->,thin] (4pt,0pt)--(4pt,2pt);
 \node at (9pt,0pt) {$#1$};
\end{tikzpicture}
}}

\def\wline#1{
\scalebox{0.7}{
\begin{tikzpicture}[baseline=-0.65ex,scale=1.3]
 \draw (0,-10pt)--(0,10pt);
 \draw [->,thin] (0,0)--(0,2pt);
 \node at (4pt,-4pt) {$#1$};
\end{tikzpicture}
}}

\def\vdashed{
    \scalebox{0.7}{
        \begin{tikzpicture}[baseline=-0.65ex,scale=1.0]
            \draw [dashed,thin] (0,-10pt)--(0,10pt);
        \end{tikzpicture}
    }
}

    \def\bra#1{\left<#1\right|}
    \def\ket#1{\left|#1\right>}
    \def\dbra#1{\left<\!\left<#1\right|\right.}
    \def\dket#1{\left.\left|#1\right>\!\right>}
    \def\braket#1#2{\left<\!#1\middle|#2\!\right>}
    \def\dbraket#1#2{\left<\!\left<#1\middle|#2\right>\!\right>}
    \def\cvectorFour#1#2#3#4{\left(\begin{array}{c}
        #1 \\ #2 \\ #3 \\ #4
    \end{array}\right)}

    \def\cvectorTwo#1#2{\left(\begin{array}{c}
        #1 \\ #2 
    \end{array}\right)}
    
    \def\rvectorTwo#1#2{\left(\begin{array}{cc}
        #1 & #2 
    \end{array}\right)}

    \def\matrixTwo#1#2#3#4{\left(\begin{array}{cc}
        #1 & #2 \\ #3 & #4
    \end{array}\right)}

    \newcommand{\news}{\setcounter{equation}{0}}
    \def\bea{\begin{eqnarray}}
    \def\eea{\end{eqnarray}}
    \numberwithin{equation}{section}
    \makeatletter
    \renewcommand*\env@matrix[1][\arraystretch]{
      \edef\arraystretch{#1}
      \hskip -\arraycolsep
      \let\@ifnextchar\new@ifnextchar
      \array{*\c@MaxMatrixCols c}}
    \makeatother
    \usepackage{color}
    
\begin{document}
    
    \title{\vskip -65pt
    \vskip 60pt
    {\bf {\large {Ishibashi States, Topological Orders with Boundaries and \\ Topological Entanglement Entropy II - Cutting through the boundary}}}\\[20pt]}
    \author{ {Ce Shen, Jiaqi Lou, Ling-Yan Hung}\\[20pt]
    $^1$State Key Laboratory of Surface Physics\\ 
    Fudan University,\\
    200433 Shanghai, China\\
    $^2$Collaborative Innovation Center of Advanced Microstructures,\\
    210093 Nanjing, China\\
    $^3$Department of Physics and Center for Field Theory and Particle Physics,\\ Fudan University\\
    200433 Shanghai, China\\
    $^4$Institute for Nanoelectronic Devices and Quantum computing,\\ Fudan University,\\ 200433 Shanghai , China\\
    }
    
    \date{\today}
    \maketitle
    \vskip 20pt
    
    \begin{abstract}
    We compute the entanglement entropy in a 2+1 dimensional topological order in the presence of gapped boundaries. Specifically, we consider entanglement cuts that cut through the boundaries. We argue that based on general considerations of the bulk-boundary correspondence,  the ``twisted characters'' feature in the Renyi entropy, and the topological entanglement entropy is controlled by a ``half-linking number'' in direct analogy to the role played by the S-modular matrix in the absence of boundaries. 
   We also construct a class of boundary states based on the half-linking numbers that provides a ``closed-string'' picture complementing an ``open-string'' computation of the entanglement entropy.  These boundary states do not correspond to diagonal RCFT's in general.  
 These are illustrated in specific Abelian Chern-Simons theories with appropriate boundary conditions.

    \end{abstract}
    
    \vfill
    
    
    \section{Introduction}\news
    
    There has been a lot of recent work exploring the effect topological boundaries have on the entanglement entropy of topological phases of matter \cite{Wang:2018edf, Shi:2018krj, chen_entanglement_2018, Shi:2018bfb,lou2019ishibashi, Hu:2019bec, Luo:2018yqb}. In our last paper \cite{lou2019ishibashi} we computed topological entanglement entropies of 2+1 dimensional topological order using  Ishibashi states at the entanglement cut. This paper is a companion paper where we inspect the situation where the entanglement cut touches the topological boundaries. 
  This case has been inspected in our previous studies based on lattice models \cite{chen_entanglement_2018}, based on recent lattice contructions of topological boundaries \cite{beigi_quantum_2011, cong_topological_2016,Hu:2017faw,Hu:2017btw,Wang:2018qvd}.  It is noticed then that there are non-universal contribution to the topological entanglement at the junction between the entanglement cut and the gapped boundary. 
 Moreover, the basis states naturally selected in the lattice model construction is naturally different from the Ishibashi state construction. In this paper, we argue that based on general considerations of bulk-boundary correspondence, the twisted characters and its modular properties studied also in the CFT literature \cite{petkova_generalised_2001}  should control the topological entanglement entropy.  The modular transformation of these twisted characters are determined by what we now understand to be the ``half-linking'' numbers $\gamma_{xc}$. Half-linking numbers are shown to diagonalize the fusion rules of defects \cite{shen2019defect}. Here, we argue that {\bf the topological entanglement entropy is dictated by these $\gamma_{xc}$ when the entanglement cut touches a boundary, in analogy to the role of the modular $S$ matrix in the absence of boundaries}\cite{dong_topological_2008,fliss_interface_2017,wen_edge_2016}.  
 We will illustrate these ideas in explicit models based on Abelian Chern-Simons theories. One very important observation, is that the precise treatment at the junction does affect the value of the topological entanglement entropy, as mentioned above. Particularly, we found that there are classes of definition of the twisted characters where the overall normalization of the half-linking number could be altered, giving up unitarity of the matrix. However, there are natural boundary conditions that recover the unitarity of the half-linking matrix. 
  
Our paper is organized as follows. In section 2, we will give general arguments that show how twisted character should feature in the computation of the entanglement entropy of a strip parallel to the axis of a cylinder with two gapped boundaries. 
We construct the appropriate ``Cardy-like - closed string states'' that recover the twisted character  obtained from edge modes having open-boundaries. In Section 3 we illustrate these ideas by computing the entanglement entropy  in the case of Abelian Chern-Simons theories.  This is computed by quantizing edge modes along the entanglement cut. With appropriate boundary conditions, they recover the twisted character.

We will briefly conclude in section 4. Some details about Abelian Chern-Simons theories are relegated to the appendix \ref{CSreview}. For a review of useful facts about gapped boundaries and anyon condensation relevant for the current paper, we refer readers to part I of our paper \cite{lou2019ishibashi}.

 \section{Entanglement cut across a gapped boundary and twisted character}

    In this section, we would like to consider entangling surfaces that touches or cut through physical boundaries and interfaces. The latter can be obtained from the former via the folding trick. We will therefore focus on the first case. 
To be concrete, we will consider the entanglement entropy of a strip connecting the top and bottom boundary of a cylinder in Abelian CS theories.

\begin{figure}
        \centering
        \begin{subfigure}[t]{0.35\textwidth}
            \includegraphics[width=\textwidth]{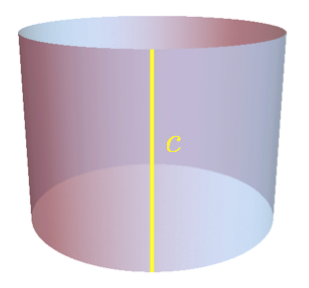}
            \caption{Shared condensed anyon basis.}
        \end{subfigure}
        \begin{subfigure}[t]{0.35\textwidth}
            \includegraphics[width=\textwidth]{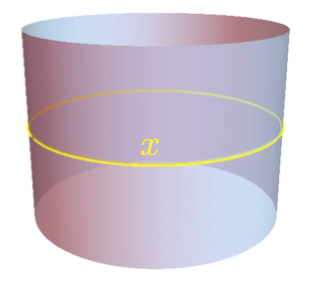}
            \caption{Confined anyon basis.}
        \end{subfigure}
        \caption{Ground state basis states on a cylinder. }
        \label{fig:cx}
        \end{figure}

Gapped boundaries are characterized by a Lagrangian algebra $\mathcal{A}$ of a modular tensor category $C$ that describes the bulk topological order in 2+1 dimensions. In physical terms, the Lagrangian algebra $\mathcal{A}$ corresponds to a set of anyons that condense at the boundary -- they are not conserved across the boundary. 
This is reviewed in detail in section 2 of our companion paper \cite{lou2019ishibashi}. We pick out a few important points here. 
Consider a cylinder with two boundaries each characterized by an anyon condensate. It generically admits non-trivial ground state degeneracy. There are two ways to construct ground state basis states. One could either consider shared condensed anyon lines connecting the two condensates at the two boundaries. An orthogonal basis carries anyon line that is confined wrt to both condensates that winds the non-contractible cycle of the cylinder. This is illustrated in figure \ref{fig:cx}.

\subsection{Twisted character in the ``open string'' frame}

Now consider an entanglement cut that touches the gapped boundary. One could compute the entanglement entropy by taking the entanglement cut as a physical boundary, and determine the Hilbert space on the two sides of the cut. Then we glue the cut back together by constructing an Ishibashi state, corresponding to the fact that the cut is fake and the theory is topological, so that the cut can be arbitrarily deformed. This is the strategy advocated in \cite{wen_edge_2016} and adopted also in our previous paper. 

In the current situation, the added complication is the physical boundaries that the entanglement cut ends on. There is some appropriate boundary condition at each end, and they are precisely conformal boundary conditions \cite{lou2019ishibashi}. Now for a given pair of boundary condition $\{x,y\}$, it determines the Hilbert space $H_{xy}$. 
Like usual open CFT, the Hilbert space $H_{xy}$ admits a decomposition
\be
H_{xy} = \oplus_z n_{xy}^z H_z,
\ee
where $H_z$ corresponds to (left-right) Virasoro representations labeled by $z$. 

Then the appropriate Ishibashi state that glues the entanglement cut back together should take the form
\be
| \Psi\rangle\rangle =\sqrt{ \mathcal{N}} \sum_{k_L, k_R} \exp(-\frac{ 2\pi \epsilon}{l} H)| z, k_L, k_R \rangle_1 \otimes |z, k_L, k_R \rangle_2 ,
\ee
where $H$ is the Hamiltonian, and $\epsilon \to 0$ is a regularization parameter, and the product of states is taken between edge states on either side of the entanglement cut. 

The trace of the $n$-th-power of the reduced density matrix $\rho$ after tracing out say edge modes labeled $2$, is then given by
\be
\textrm{tr} \rho^n = \mathcal{N}^n \chi_z (q, \bar{q}), \qquad q = \exp(i\tau), \qquad \tau = \frac{ i 2\pi \epsilon}{l}.
\ee
Recall that a Lagrangian algebra $\mathcal{A}$ is in one-to-one correspondence with modular invariants. Each condensed sector $c_i \in \mathcal{A} $ labels a pair of left-right Virasoro representation that features in the modular invariant.  i.e. 
\be
c_i \equiv \{h_{L_i}, h_{R_i}\}, \qquad \exp( i \theta_{c_i}) = \exp(2\pi i (h_L- h_R)),
\ee
where $\{h_{L_i}, h_{R_i}\}$ are the conformal dimensions of the primariy operator corresponding to the condensed sector and $\theta_{c_i}$ is the topological spin of the sector. 

The crucial observation here is that these  $x,y,z$ are all labels of line operators in the modular invariant CFT defined by the Lagrangian algebra $\mathcal{A}$ when the two boundaries share the same $\mathcal{A}$. From the perspective of the topological order, these line operators are the {\bf confined sectors} in the boundary condensate $\mathcal{A}$. 
The $\chi_z$ are twisted characters corresponding to the insertion of these line operators. Some special cases of these line operators and the corresponding twisted characters have been discussed in the CFT literature \cite{petkova_generalised_2001}.  
\subsection{Line operators and twisted character in the ``closed string'' frame}

The twisted character $\chi_z$ satisfies
\be \label{X_x}
\sum_{z} n_{xy}^z \chi_z(\tilde q, \tilde{\bar{q}}) = \textrm{tr}_{x|y} (X_x X_y q^{-L_0} \bar{q}^{-\bar{L}_0}), \qquad \tilde q \equiv q^{-1/\tau}.
\ee
where $X_{x,y}$ are line operators, and 
\be
\chi_z(\tilde q, \tilde{\bar{q}}) = \sum_i \gamma_{z c_{i}} \chi_{c_i}(q, \bar{q}),
\ee
where $\gamma_{z c_i}$ coincides with the half-linking matrix discussed in \cite{shen2019defect} and reviewed in the appendix. 

The line operators can be written explicitly in the following form
  \begin{eqnarray}
        \label{eq:em_Cardy}
       X_x =\sum_{c_i \in\mathcal{A}}^{}\frac{\gamma_{xc_i }}{\sqrt{\gamma_{0c_i}}} \sum_{k_L, k_R} | c_i , k_L, k_R \rangle \langle c_i, k_L, k_R |,
    \end{eqnarray}
where $|c_i, k_L, k_R\rangle $ are generic descendants at level $\{k_L, k_R\}$ of the left-right primary $c_i$ inside the Hilbert space defined by the modular invariant characterized by the Lagrangian algebra $\mathcal{A}$. 
These are analogues of boundary states satisfying the Cardy conditions. Their construction in the case of diagonal RCFTs was discussed in \cite{petkova_generalised_2001}. Here, we give a general form using data of the bulk topological order and the half linking matrix $\gamma$.

These line operators $X_x$ can in fact be re-arranged into a pair of conformal boundary states via the folding trick, by reversing the bra $\langle c_i, k_L, k_R|$ into a ket and turning a left moving mode into a right moving mode and vice versa. i.e. 
\be
\sum_{k_L,k_R} |c_i, k_L ,k_R \rangle \langle c_i, k_L, k_R|  \to  \sum_{k_L} |c_i, k_L \rangle |c_i , k'_R = k_L \rangle   \otimes  \sum_{k_R} |c_i, k_R \rangle |c_i , k'_L = k_R \rangle = |c_i\rangle\rangle_L \otimes |c_i \rangle\rangle_R , 
\ee
where $|c_i \rangle \rangle$ is a conformal boundary state. These conformal boundary states are precisely those basis conformal boundary states that we constructed to describe a condensed anyon ending at the gapped boundary. (See equation (2.15) in the Abelian case and (2.49) for non Abelian ones in part I of our paper \cite{lou2019ishibashi}. ) 
 i.e. We can define in place of $X_x$,
\be
|B_{x}\rangle = \sum_{c_i \in\mathcal{A}}^{}\frac{\gamma_{xc_i }}{\sqrt{\gamma_{0c_i}}} |c_i \rangle \rangle_L \otimes |c_i \rangle \rangle_R. 
\ee
Then, equation (\ref{X_x}) can be re-arranged as overlaps of boundary states $|B_x\rangle$. 
Their overlaps recover the twisted characters considered above using the analogue of the Cardy condition:
\begin{eqnarray}
\bra{B_z}e^{-H/\delta}\ket{B_y}&=&\sum_{c\in\mathcal{C}}\frac{\gamma_{cz}^{\dagger}\gamma_{yc}}{\gamma_{0c}}\chi_c(\tilde{q})\nonumber\\
&=&\sum_{c,x}\frac{\gamma_{cz}^{\dagger}\gamma_{yc}\gamma_{cx}^{\dagger}}{\gamma_{0c}}\chi_{x}(q)\nonumber\\
&=&\sum_{x}n_{xz}^{y}\chi_{x}(q)
\end{eqnarray}
where $n_{xz}^{y}$ is the fusion coefficient of the confined sectors -- or equivalently, the line operators of the CFT. We have made use of the defect version of the Verlinde formula found in \cite{shen2019defect}. We use little $n$ to distinguish the fusion of confined sectors from that of the bulk anyons $N_{ab}^c$.

Note that when the two edges of the cylinder are characterized by different anyon condensates, then these line operators $X_a$ are interface operators between two different modular invariant CFT's, each determined by the Lagrangian algebra at each end of the cylinder. These twisted characters then transform under modular transformation generically as
\be
\chi_z (\tilde q)  =\sum_{c_i \in \mathcal{A}_\mu \cap \mathcal{A}_\nu}  \gamma^{(\mu|\nu)}_{z c_i} \chi_{c_i}(q).
\ee

\subsection{Entanglement entropy from twisted characters}

The computation of the entanglement entropy of this ground state eigen-basis is thus given by
\begin{align} \label{eq:topEE}
&S^{(\mu|\nu)}_{EE}(N \,\,\textrm{strips on a cylinder}) = 2N\left(\lim_{\substack{n\to 1\\ \epsilon \to 0}} \frac{1}{n-1} \ln \frac{\chi_x(\exp(-n \epsilon/l))}{\chi_x(\exp(-\epsilon/l))^n} \right)  \\
&\approx N \left( \lim_{\substack{n\to 1\\ \epsilon \to 0}} \frac{1}{n-1}\left(\ln \frac{\chi_{c_{\textrm{min}}}(\exp(- l/(n \epsilon))) }{\chi^n_{c_{\textrm{min}}}(\exp(-l/\epsilon))}\right) - \ln \gamma^{(\mu|\nu)}_{x_{c_{\textrm{min}}}}\right)
\end{align}
where $\chi_x$ are the characters of the Virasoro representation $x$.


The labels $c_{\textrm{min}}$ denotes the primary corresponding to the shared condensed sectors between boundary $\mu$ and $\nu$,  whose conformal dimension is the smallest. 

We will illustrate the above in Abelian Chern-Simons theory in the next section. We note that in the case corresponding to $Z_N$ gauge theories, where  the electric condensate resides at one boundary and the magnetic condensate the other, we found that the expansion of the character $\chi_{c_{\textrm{min}}}$ led to a factor of $\sqrt{2}$ following from a non-trivial choice of normalization that records the majorana zero mode, even though $\gamma^{(E|M)}_{xc} = 1$ there \cite{shen2019defect}.

In the simpler case where both boundaries the entanglement cut touches carry the same set of condensed anyons, $c_{\textrm{min}} = 0$ i.e. $\chi_{c_{\textrm{min}}}$ is the vacuum Virasoro character, and the above expression reduces to
\be
S^{\textrm{same bc}}_{EE}(N \,\,\textrm{strips on a cylinder}) = 2N (\frac{c l}{12 \epsilon} - \ln \gamma_{x0}).  
\ee
 Where the area term follows from expanding the vacuum character in the $\epsilon \to 0$ limit, and the topological entanglement entropy is given simply by $\ln \gamma_{x0}$, which plays the same role as the modular matrix $S$ in the absence of boundaries. 
 
 Consider a bulk phase $B = C\boxtimes \bar{C}$, where $\bar{C}$ denotes the time-reversal of $C$,  and that the boundaries are characterized by a ``diagonal'' Lagrangian algebra composed of anyons $\{\sigma \boxtimes \bar{\sigma}\}$ for all $\sigma \in C$.  In this case, the cylinder can be unfolded into a torus made of $C$. The entanglement cut can be unfolded into a circle. In this case the confined anyons and condensed anyons can both be labelled by objects in $C$, and it is found that $\gamma_{xc} = S^C_{xc}$\cite{shen2019defect}. Our result then reduces to the case of entanglement entropy of $C$ on a closed surface, as expected. 
 
 Generic linear combinations of the $x$ eigen-basis leads to a reduced density matrix with superselection sectors labelled by $x$'s. The probability $p_x$ of each sector thus generate an extra classical piece in the entanglement entropy $\Delta S_{EE} = \sum_x -p_x\ln p_x$. 
 

    \section{Examples in Abelian Chern-Simons theories}
    
       We consider the case in which a cylinder with gapped boundaries is bi-partitioned into strip regions $R$ and $\bar{R}$. The bulk theory considered here is again the $\mathbb{Z}_N$ toric code with action (\ref{CSaction}) and $K$ matrix (\ref{kmatrixZN}).
        The entanglement entropy between region $R$ and $\bar{R}$ is calculated for various boundary conditions in the following.
        To make things explicit, we will in particular consider the top and bottom physical boundaries $B_{1,2}$ to take the electric/magnetic boundary conditions in turn. 
        Along each of the entanglement cut, we will again recover a pair of edge CFT. For reasons that will become clear later on, we will denote the edge modes on the entanglement cut by $\phi^{l_{1,2}, r_{1,2}}_{I \in \{1,2\}}$. The superscripts $l_{1,2}, r_{1,2}$ refer to the modes on the pair of vertical entanglement cut $b_{1,2}$. Specifically,  $l_i$ are the degrees of freedom on the left edge of the cut, while $r_i$ dofs  on the right edge of the cut. 
        
        The vertical cut is of length $l$. We will for now work with dimensionless coordinates, and scale it to $\pi$. We will restore dimensions at the end. 
        The physical boundaries $B_{1,2}$ are now located at $x=0,\pi$ respectively. The condensate dictates specific boundary condition that we should impose on the modes $\phi^{l_i,r_i}_{I}$ at these end points. 
        
        In equations (\ref{celectric}) and (\ref{cmagnetic}) we presented the conformal boundary conditions that should be satisfied by the edge modes located at the physical boundary,  appropriate for the electric and magnetic boundary condensate respectively. 
        In the current situation, these physical boundaries are the end points of the entanglement cut. Therefore the appropriate boundary condition should be replaced by
        \be
        \label{eq:em_bc}
    \sum_J l^{e,m}_J \partial_t \phi^{l_i, r_i}_J\vert_{x=0} =   \frac{1}{2\pi}\partial_t \phi^{l_i, r_i}_{I} \vert_{x={0}} =0,
        \ee
        where $l^{e,m}$ are the charge vectors of the condensed anyon characterizing the boundary as reviewed in section \ref{app:revgapped}. Specifically, here we have $I=1$ for an electric boundary condition and $I=2$ for a magnetic boundary condition. These conditions are satisfied by both the modes on the left/right of the entanglement cut. Similarly boundary condition has to be imposed at $x=\pi$. The problem thus reduces to one of quantizing an ``open-string'' with specific boundary conditions at the end points. 
        We will consider combinations of these boundary conditions in turn. 
        From this discussion, it should be clear that the modes that we obtained by quantizing the scalar field with boundary conditions (\ref{eq:em_bc}) correspond to ``boundary-changing operator" (bco) that connects the boundaries at the two ends. The topological entanglement entropy then follows from the modular properties of the characters of these bcos. 
        
        \begin{figure}
        \centering
        \begin{subfigure}[t]{0.35\textwidth}
            \includegraphics[width=\textwidth]{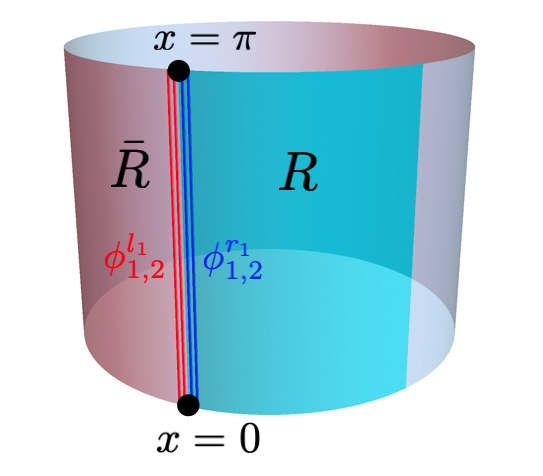}
            \caption{folded picture}
            \label{fig:embdy}
        \end{subfigure}
        ~ 
          \begin{subfigure}[t]{0.5\textwidth}
            \includegraphics[width=\textwidth]{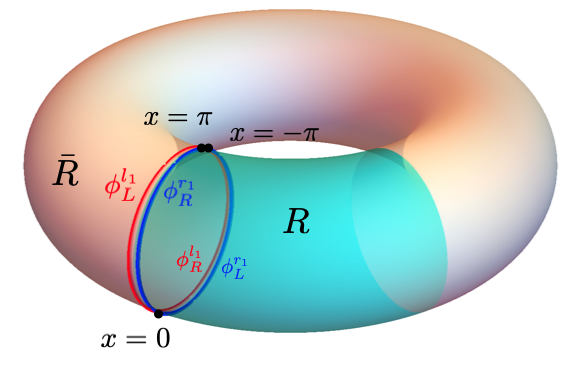}
            \caption{unfolded picture}
            \label{fig:torus}
        \end{subfigure}
        
        \label{fig:zmzn}
        \caption{``Open string'' modes in the folded picture and ``closed string'' modes in the unfolded picture.}
        \end{figure}

    \subsubsection{electric + magnetic b.c.} \label{sec:EM}
       Consider the case where the entanglement cut ends on two different boundaries. We will illustrate this case in detail.  
              In this case the upper $S^1$ edge ($x=\pi$) is electric while the lower $S^1$ edge ($x=0$) is magnetic. 
              From the discussion in \cite{cong_topological_2016}, we recall that the ground state here is unique. 
              The boundary conditions on the edge modes are
              \begin{eqnarray}
                  \begin{cases}
                  \partial_t\phi_2^{l_{1,2}, r_{1,2}}|_{x=0}=0 \\
                  \partial_t\phi_1^{l_{1,2}, r_{1,2}}|_{x=\pi}=0
                  \end{cases}.
              \end{eqnarray}
              In terms of the left and right moving chiral fields as in (\ref{left_rightmodes}), we have
              \begin{eqnarray}
              \begin{cases}
              \partial_t\phi_L^{l_{1,2}, r_{1,2}}|_{x=0}- \partial_t \phi_R^{l_{1,2}, r_{1,2}}|_{x=0} = 0\\
              \partial_t\phi_L^{l_{1,2}, r_{1,2}}|_{x=\pi}+\partial_t\phi_R^{l_{1,2}, r_{1,2}}|_{x=\pi} = 0
              \end{cases}.
              \end{eqnarray}
              
              Since $\phi_R(x)$ lives in the interval $(0,\pi)$ and  $\phi_R(-x)$ lives in the interval $(-\pi,0)$, we can then simplify notations by combining  $\phi^{l_{1,2}, r_{1,2}}_{L,R}$  into  ``closed-string'' fields defined on $(-\pi,0)\cup(0,\pi)$. This gives
              \begin{eqnarray}
                  \label{eq:united_mode}
              \begin{cases}
                  \Phi^{1,2}_L(x)=\phi^{l_{1,2}}_L(x) \oplus \phi^{l_{1,2}}_R(-x) \\
                  \Phi^{1,2}_R(x) =\phi^{r_{1,2}}_L(x) \oplus \phi^{r_{1,2}}_R (-x)
              \end{cases},
              \end{eqnarray}
              The boundary conditions above translate into continuity and anti-periodicity relations of $\Phi_L$
              \begin{eqnarray}
                  \begin{cases}
                  \Phi^{1,2}_L(x=0^{-})=\Phi^{1,2}_L(x=0^{+}) \\
                  \Phi^{1,2}_L(x=-\pi)=-\Phi^{1,2}_L(x=\pi). \\
                  \end{cases}.
              \end{eqnarray}
              
              Applying the anti-periodicity condition to a general expansion of the chiral mode we obtain
              \begin{eqnarray}
                  \Phi^{1,2}_L=\Phi^{1,2}_{0L}+ 2\pi P^{1,2}_L x+i\sum_{n\ne 0}^{}\frac{\alpha^{1,2}_{L,n}}{n}e^{-inx},
              \end{eqnarray}
              one sees that 
              \be
              \Phi^{1,2}_{0L}=0, \qquad P^{1,2}_L=0.
              \ee
              This corresponds to the fact that we are working with an eigenstate with trivial flux crossing the entanglement cut. This is the only shared confined sector between an electric condensate and a magnetic one. 
               
              Also $n$ is a half-integer rather than an integer. Finally the mode expansion becomes 
              \be
              \Phi^{1,2}_L=i\sum_{n\in\frac{1}{2}+\mathbb{Z}}^{} \frac{\alpha^{1,2}_{L,n}}{n}e^{-inx}.
              \ee
              The field $\Phi^{1,2}_R$ satisfies exactly the same set of equations, except where we have to take $x\to -x$ and replace $\alpha_L$ by $\alpha_R$.
              
              Each entanglement cut then corresponds to a generalized Ishibashi state matching $\Phi^{i}_{L}$ and $\Phi^i_R$, in a way completely analogous to the boundary Ishibashi state in \cite{lou2019ishibashi} except that we have now only half the number of modes. 
              The normalized Ishibashi state at each cut becomes
              \begin{eqnarray}
              \dket{0}_{b_i}={e^{- \frac{2\pi}{2l}\epsilon H_i}}   \exp\left(\sum_{m\in\frac{1}{2}+\mathbb{N}}^{}\frac{1}{m} \alpha^i_{-m}\bar{\alpha}^i_{-m} \right)  \ket{0}_{b_i},
              \end{eqnarray} 
              where the Hamiltonian $H_i=(L^i_0+\bar{L}^i_0-1/12)$, $i$ again denoting the $i-$th entanglement cut. The Hamiltonian is inserted as a UV regularization \cite{wen_edge_2016}, with the UV cutoff scale given by $\epsilon$.
              Note that we have restored the dimensionful parameter $l$, the length of the cut, in the expression.
              There is an extra factor $2$ in $2l$ since the length of a closed string on the circle doubles that of an open string on the vertical cut.

              The normalization constant $\mathcal{N}_0$ can be determined by $\dbraket{0}{0}=1$,
              \begin{eqnarray}
                  \label{eq:em_char}
                  \mathcal{N}_0&=&\bra{0}\exp\left(\sum_{n\in\frac{1}{2}+\mathbb{N}}^{}\frac{1}{n} \alpha_{-n}\bar{\alpha}_{-n} \right) e^{-\frac{4\pi\epsilon H}{2l}}\exp\left(\sum_{m\in\frac{1}{2}+\mathbb{N}}^{}\frac{1}{m} \alpha_{-m}\bar{\alpha}_{-m} \right)\ket{0} \nonumber\\ 
                  &=&q^{-\frac{1}{24}}  \sum_{m\in\frac{1}{2}\mathbb{N}}^{} q^m \bar{p}(m)  \nonumber\\
                  &=&q^{-\frac{1}{24}} \prod_{r\in\frac{1}{2}+\mathbb{N}}^{} \frac{1}{1-q^r} \nonumber\\ 
                  &=&q^{-\frac{1}{16}} \sqrt{\frac{\eta(q)}{\theta_4(q)}} 
              \end{eqnarray}
              where $q=e^{\frac{-8\pi\epsilon}{2l}}$ and the generating function for degeneracy $\bar{p}(m)$ is given by:
              
              \begin{eqnarray}
              \sum_{m\in\frac{\mathbb{N}}{2}}^{}\bar{p}(m)x^m=\prod_{r\in\frac{1}{2}+\mathbb{Z}}^{}\frac{1}{1-x^r}
              \end{eqnarray}

              For the trivial topological sector, the lowest eigenstates of $L_0$ are listed below. 
              
              \begin{tabular}{ccl}
                  \hline 
                  level $m$ & degeneracy $\bar{p}(m)$ & states \\ \hline
                  0 & 1 & $\ket{0}$ \\ 
                  $\frac{1}{2}$ & 1 & $\alpha_{-\frac{1}{2}}\ket{0}$ \\
                  1 & 1 & $\alpha_{-\frac{1}{2}}^2\ket{0}$ \\
                  $\frac{3}{2}$ & 2 & $\alpha_{-\frac{1}{2}}^3\ket{0}$, $\alpha_{-\frac{3}{2}}\ket{0}$ \\
                  \hline 
                  
              \end{tabular}
              
              Now putting together the two entanglement cuts,  the state may be written as a direct product of the two boundaries.
              \begin{eqnarray*}
                  \dket{0}&=&\dket{0}_{b_1}\otimes\dket{0}_{b_2} \nonumber \\
                  &=&{e^{-\frac{2\pi\epsilon H_{1}}{2l}}}  \exp\left(\sum_{m\in\frac{1}{2}+\mathbb{N}}^{}\frac{1}{m} \alpha_{-m}^{1}\bar{\alpha}_{-m}^{1} \right)  \ket{0}_{b_1} \otimes {e^{-\frac{2\pi\epsilon H_{2}}{2l}}} \exp\left(\sum_{n\in\frac{1}{2}+\mathbb{N}}^{}\frac{1}{n} \alpha_{-n}^{2}\bar{\alpha}_{-n}^{2} \right)  \ket{0}_{b_2} 
              \end{eqnarray*} 
              
              The reduced density matrix $\rho_L$ is obtained by tracing out the anti-chiral part of the full density matrix $\rho= \mathcal{N}_0^{-2} \dket{0}\dbra{0}$. 
              We therefore have
              \begin{eqnarray}
                  \textrm{Tr}_L\rho_L^n&=&\left( \frac{\mathcal{N}_0(q^n)}{\mathcal{N}_0(q)^n} \right)^{(a)} \left( \frac{\mathcal{N}_0(q^n)}{\mathcal{N}_0(q)^n} \right)^{(b)}  \nonumber \\
                  &=& \left(\left(\sqrt{{\frac{\theta_4(q)}{\eta(q)}}}\right)^{n}  \sqrt{\frac{\eta(q^n)}{\theta_4(q^n)}} \right)^2 \nonumber \\
                  &=& \left(\left(\sqrt{{\frac{\theta_2(\tilde{q})}{\eta(\tilde{q})}}}\right)^n \sqrt{ \frac{\eta(\tilde{q}^{1/n})}{\theta_2(\tilde{q}^{1/n})}}\right)^2 \nonumber \\
                  &\overset{l/\epsilon\rightarrow\infty}{\longrightarrow}&2^{n-1}\tilde{q}^{\frac{1}{12}(n-\frac{1}{n})} 
              \end{eqnarray}
              where in the last step the thermodynamic limit($l/{\epsilon}\rightarrow\infty$) is taken, and the entanglement entropy is
              \begin{eqnarray}
                  S=\lim\limits_{n\rightarrow1}\frac{1}{n}\log \textrm{Tr}_L\rho_L^n = 2(\frac{\pi l}{12\epsilon}-\log\sqrt{2}).
              \end{eqnarray}
              We make it explicit with the overall factor of 2,  that the two entanglement cuts of the strip contribute equally to the entanglement entropy. 
              We comment here that the non-trivial topological entanglement entropy $-\log\sqrt{ 2}$ comes from the two ground states of the orbifold. This is 
              in fact the Majorana mode supposedly trapped at the junction between the electric and magnetic boundaries \cite{Barkeshli:2013yta,cong_topological_2016, lan_gapped_2015,shen2019defect}.

    \subsubsection{electric + electric b.c.} \label{sec:EE}
In this case the boundary condition becomes
  \begin{eqnarray}
        \begin{cases}
        \partial_t\phi_1^{l_{1,2}, r_{1,2}}|_{x=0}=0 \\
        \partial_t\phi_1^{l_{1,2}, r_{1,2}}|_{x=\pi}=0
        \end{cases}.
    \end{eqnarray}
    The analysis is very similar to the previous case, and we will only outline the procedure. 
  The allowed set of zero modes here is given by
    \be \label{eq:ee_zero_mode}
    P^1=0, \qquad  P^2= N a + b,
    \ee
for all $a\in \bZ$, and each $b$ satisfying $0\leq b\leq N-1$ parametrizes an independent sector. 
For each fixed $b$, this is an eigenstate of definite flux crossing the entanglement cut. In particular, we note that these are distinct confined sectors relative to the electric condensate at the boundaries -- i.e. they are the magnetic charges.
Ishibashi states can be constructed for each fixed $b$ that entangles the $l,r$ modes at each cut exactly as in the previous subsection, except that the  oscillatory modes $\alpha_n$ now take integer values of moding $n$.  
We note that the sum over zero modes contain only a sum over $a$. 
Alternatively, from the perspective of a combined mode $\Phi_{L,R}$ constructed as in (\ref{eq:united_mode}), it is describable by an effective action characterized by $1\times1$ $K$-matrix $K_{eff}=(N)$, since the number of edge modes is halved in the presence of boundaries.
  Then the effective quantum dimension at each entanglement cut is given by $D_{eff}=\sqrt{N}=\sqrt{D}$, and each entanglement cut contributes $-\ln D_{eff}=-1/2\ln N$ to the entanglement entropy. (Compare with (\cite{lou2019ishibashi}) where each bulk cut contributes $-\ln N$.) Combining the contribution of the two cuts we recover the results in \cite{chen_entanglement_2018}.

    \subsubsection{magnetic + magnetic b.c.} \label{sec:MM}
   The appropriate boundary conditions in this case are given by
 \begin{eqnarray}
        \begin{cases}
        \partial_t\phi_2^{l_{1,2}, r_{1,2}}|_{x=0}=0 \\
        \partial_t\phi_2^{l_{1,2}, r_{1,2}}|_{x=\pi}=0
        \end{cases}.
    \end{eqnarray}
    This leads to the following allowed set of zero modes
     \be \label{eq:mm_zero_mode}
    P^2=0, \qquad  P^1= N a + b. 
    \ee
This analysis of this case is equivalent to the electric + electric case. The result of the entanglement entropy is identical. 
We emphasize again that a natural basis Ishibashi state parametrized by a fixed $b$ corresponds to distinct confined sectors relative to the boundary condensates.

\subsection{The ``closed-string'' frame}

       Written explicitly, the ``closed string'' electric and magnetic Ishibashi states are respectively given by 
    \begin{eqnarray}
        \dket{a}_{E}=\sum_{b\in\mathbb{Z}}^{}\exp \left(-\sum_{l=1}^{\infty}\frac{1}{l}\alpha_{-l}\bar{\alpha}_{-l}\right) \ket{P_L=P_R=\frac{a+bN}{\sqrt{N}}}\\
        \dket{a}_{M}=\sum_{b\in\mathbb{Z}}^{}\exp \left(\sum_{l=1}^{\infty}\frac{1}{l}\alpha_{-l}\bar{\alpha}_{-l}\right) \ket{P_L=-P_R=\frac{a+bN}{\sqrt{N}}}
    \end{eqnarray}
    where the primary states are orthonormal 
    $ \langle P_L,P_R|P_L^{\prime},P_R^{\prime}\rangle=\delta_{P_L P_L^{\prime}}\delta_{P_R P_R^{\prime}}$.
    
   In Abelian Chern-Simons theory for example it satisfies 
    \be \label{eq:cstate}
    l_I \partial_x \phi_I \dket{c} = 0, \qquad l \in L,
    \ee
where $L$ is the set of charge vectors characterizing the gapped boundary condensate as reviewed in (\ref{eq:condensed_vecs}). 
The subscript $x$ of $\gamma_{xc}$ (as opposed to the spatial coordinate $x$ appearing in $\partial_x$ in equation (\ref{eq:cstate})) denotes a confined anyon.  As demonstrated in the examples above in sections \ref{sec:EM}, \ref{sec:EE} and \ref{sec:MM}, it is natural to construct eigenbasis so that there is definite anyon flux crossing the entanglement cut. The distinct states are thus labeled by sectors confined wrt both condensates at the boundaries of the cylinder. The boundary states in the ``closed string'' channel are therefore naturally labelled by the same basis. The corresponding ``closed-string'' boundary states that recover the ``open-string'' picture satisfying (\ref{eq:cstate}) is natural in the sense that to convert between the annulus and the cylinder, it is as if we are exchanging $x$ and $t$, and thus replacing the boundary condition (\ref{eq:em_bc}) by (\ref{eq:cstate}).
We note that in the current example,  
\be
\gamma_{xc}=\frac{1}{\sqrt{N}}\exp (-\frac{2\pi icx}{N})=\frac{S_{xc}}{\sqrt{S_{0c}}}.
\ee

    \begin{description}
    \item[In e+m case] GSD=1, no topological sectors remain confined relative to both boundaries, which means every non-contractible Wilson loop can be freely absorbed into the boundaries. Therefore, the only ``confined'' sector is trivial, $x=0$. So we can omit the confined index $x$ in this case. The overlap of electric and magnetic boundary states projects to the trivial sector:
    \begin{eqnarray}
        \label{eq:em_closed}
        {}_{E}\bra{B}e^{-H/\delta}\ket{B}_M&=&\frac{1}{\sqrt{N}}\,{}_E\!\dbra{0}e^{-H/\delta}\dket{0}_M \nonumber\\
        &=&\frac{1}{\sqrt{N}}\bra{0}\exp \left(-\sum_{l=1}^{\infty}\frac{1}{l}\alpha_{-l}\bar{\alpha}_{-l}\right) e^{-H/\delta} \exp \left(\sum_{l=1}^{\infty}\frac{1}{l}\alpha_{-l}\bar{\alpha}_{-l}\right) \ket{0} \nonumber\\
        &=&\frac{e^{\frac{\pi}{6\delta}}}{\sqrt{N}}\prod_{k=1}^{\infty}\frac{1}{1+e^{-4\pi k/\delta}}\nonumber\\
        &=&\frac{\sqrt{2}}{\sqrt{N}}\sqrt{\frac{\eta(\tilde{q})}{\theta_2(\tilde{q})}}
    \end{eqnarray}
    where $\ket{0}$ denotes the lowest-energy ground state $\ket{P_L=0,P_R=0}$ and $\tilde{q}=e^{-4\pi/\delta}$. Note that  the definition of the  boundary state $\ket{B}_E$ (resp. $\ket{B}_M$) involves the half-linking matrix $\gamma^{(E|E)}$ (resp. $\gamma^{(M|M)}$). 
    The  factor ${\sqrt{\frac{2}{N}}}$ here is due to normalization of Ishibashi states \cite{blumenhagen_introduction_2009} . Then the modular transformation of $\eta$ and $\theta_2$ function (see appendix \ref{app:eta_theta_function}) recovers the open string character (\ref{eq:em_char}) under the identification $\delta=2\epsilon/l$.

    \item[In e+e/m+m case] GSD=$N$, labeled by the electric Wilson lines connecting physical boundaries (condensed sector $c$), or altenatively labeled by incontractible magnetic Wilson loops (confined sector $x$). The transformation matrix relating the two bases is $\gamma$ that will appear below. The open string quantization (\ref{eq:ee_zero_mode}) gives the following character
    \begin{eqnarray}
        \label{eq:ee_open}
        \chi_x(q)&=&\sum_{b_1,b_2\in\mathbb{Z}}\bra{\frac{x+b_1N}{\sqrt{N}}}\exp \left(\pm\sum_{l=1}^{\infty}\frac{1}{l}\alpha_{-l}\bar{\alpha}_{-l}\right) e^{\frac{-2\epsilon\pi}{l} H} \exp \left(\pm\sum_{l=1}^{\infty}\frac{1}{l}\alpha_{-l}\bar{\alpha}_{-l}\right)\ket{\frac{x+b_2N}{\sqrt{N}}}
        \nonumber\\
        &=& \sum_{b\in\mathbb{Z}}^{}\frac{\exp(-\pi\delta\frac{(x+bN)^2}{2N})}{\eta(q)}=\frac{1}{\sqrt{N}\eta(\tilde{q})}\sum_{k\in\mathbb{Z}}\exp\left(-\frac{2\pi}{N\delta}k^2-\frac{2\pi ix}{N}k\right)
        \nonumber\\
        &=&\frac{1}{\sqrt{N}\eta(\tilde{q})}\sum_{n\in\mathbb{Z}}\sum_{c=0}^{N-1}\exp \left( -\frac{2\pi}{N\delta}(nN+c)^2-\frac{2\pi icx}{N} \right)
        \nonumber\\
        &=&\sum_{c=0}^{N-1}\gamma_{xc}\chi_c(\tilde{q})
    \end{eqnarray}
    where $q=e^{-{2\pi\epsilon/l}}=e^{-\pi\delta}$ and $\tilde{q}=e^{-4\pi/\delta}$ under the identification $\delta=2\epsilon/l$. Poisson resummation is performed in the second line. In the third line we rewrite $k=nN+m$ and split the sum $\sum_{k\in\mathbb{Z}}$ into a double sum $\sum_{n\in\mathbb{Z}}\sum_{c=0}^{N-1}$.  As before, $c$ labels different topological sectors condensed at the boundaries, while $n$ labels the primary states corresponding to the same topological sector.


    We focus on \textbf{e+e} case in the following, since \textbf{m+m} case is analyzed in a similar manner. Note that the building block of boundary state overlap, $\dbra{c}e^{-H/\delta}\dket{c}$, can be considered as the amplitude of a closed string propagating $1/\delta$ along the Euclidean time direction, and hence as an open string amplitude via open-closed duality. Explicitly, the building block can be expressed as
    \begin{eqnarray}
     &&  \dbra{c}e^{-H/\delta}\dket{c} \nonumber\\
        &=&\sum_{n_1,n_2\in\mathbb{Z}}\bra{\frac{c+n_1N}{\sqrt{N}}}\exp \left(\pm\sum_{l=1}^{\infty}\frac{1}{l}\alpha_{-l}\bar{\alpha}_{-l}\right) e^{-H/\delta} \exp \left(\pm\sum_{l=1}^{\infty}\frac{1}{l}\alpha_{-l}\bar{\alpha}_{-l}\right)\ket{\frac{c+n_2N}{\sqrt{N}}}\nonumber\\
        &=&\frac{1}{\eta(\tilde{q})}\sum_{n\in\mathbb{Z}}\exp\left( -\frac{4\pi}{\delta}\frac{(c+nN)^2}{2N} \right)\nonumber\\
        &=&\chi_c(\tilde{q})
    \end{eqnarray}

    
    With the building block at our disposal, the overlap between boundary Cardy states (\ref{eq:em_Cardy}) is easily identified with open string amplitude (\ref{eq:ee_open}):
    \begin{eqnarray}
        \label{eq:ee_closed}
        \bra{B_z}e^{-H/\delta}\ket{B_y}&=&\sum_{c\in\mathcal{C}}\frac{\gamma_{cz}^{\dagger}\gamma_{cy}}{\gamma_{0c}}\dbra{c}e^{-H/\delta}\dket{c}
        \nonumber\\
        &=&\sum_{c\in\mathcal{C}}\frac{\gamma_{cz}^{\dagger}\gamma_{cy}}{\gamma_{0c}}\chi_c(\tilde{q}).
    \end{eqnarray}
    More generically, the $\gamma$ matrix for the bulk topological order (\ref{kmatrixZN}) is given by 
    \begin{eqnarray}
        \gamma_{xc}=\frac{S_{xc}}{\sqrt{S_{0c}}}=\frac{1}{\sqrt{N}}\exp(2\pi i l_x^TK^{-1}l_c)
    \end{eqnarray}
    where we assign to the condensed sector $c$ a charge vector $l_c\in L$, and assign to the confined sectors charge vectors $l_{y,z}$. The designation is not unique since any charge vector $l^{\prime}=l+K\Lambda$ with $\Lambda$ an integer vector corresponds to the same topological sector.  Here, we have $l_c^T = (c,0)$, whereas $l^T_{y} = (0,y)$ and $l^T_z= (0,z)$. 
    If we choose $z-y\equiv x\quad\text{mod }N$, then ${\gamma_{zc}^{\dagger}\gamma_{yc}}/{\gamma_{0c}}=\gamma_{xc}$ and (\ref{eq:ee_closed}) is exactly equal to (\ref{eq:ee_open}).
    
%
    \end{description}

   \subsection{Generic 2-2 K matrix theories, and a condensed-confined duality}\label{sec:condensed_confined_duality}
Consider Abelian topological order given by K matrix, in order that the bulk can support gapped boundary, a necessary but not sufficient condition is that the quadratic form must have total signature $0$, namely there exists a matrix $A$ such that the K matrix can be diagonalized as follows:
\begin{equation}
    \label{eq:diagonalize}
    A^T K A = \matrixTwo{1}{0}{0}{-1}
\end{equation}

 In Abelian topological order, the gapped boundary is described by a Lagrangian subgroup $M$. $M$ is an integer matrix whose column vectors are the condensed anyons, and the group structure is provided by anyon fusion.
 For convenience we call the set of condensed anyons the \textit{condensate}, and the set of confined anyons the \textit{confinate}.
 The condensate $M$ satisfies self-null and mutual-null condition $M^T K^{-1} M=0$. So 
\begin{align} \label{eq:bc1}
    M=\{\mb{m}\in\Z^2|M^T K^{-1} \mb{m}=0 \} = \ker(M^T K^{-1}) \cap \Z^2 = K \ker(M^T) \cap \Z^2
\end{align}

Given a boundary condensate $M$, the open string touching this boundary has to satisfy the boundary condition (\ref{eq:em_bc}), which is equivalent to $M^T AA^TN=0$ where $N$ is the confinate\footnote{Here and in the following we assume that the confined sectors have bulk representatives, and we take these bulk representatives as elements of the confinate $N$.} corresponding to the condensate $M$. So
\begin{equation} \label{eq:bc2}
    N=\{\mb{n}\in\Z^2|M^T AA^T \mb{n}=0 \} = \ker(M^T AA^T) \cap \Z^2 = (AA^T)^{-1} \ker(M^T) \cap \Z^2
\end{equation}

For a cylinder topology with two physical boundaries $M_{\mu}$ and $M_{\nu}$, the set of shared condensed anyons is 
\begin{align}
    M_{\mu} \cap M_{\nu} &= \{\mb{m}\in\Z^2|\ M_{\mu}^T K^{-1} \mb{m}=0,\ M_{\nu}^T K^{-1} \mb{m}=0 \} \nonumber\\
    &=K \ker(M_{\mu}^T) \cap K \ker(M_{\nu}^T) \cap \Z^2 \nonumber\\
    &=K \left( \ker(M_{\mu}^T) \cap \ker(M_{\nu}^T) \right) \cap \Z^2 \nonumber\\
    &= K \ker(M_{\mu}\vdashed M_{\nu})^T \cap \Z^2 \nonumber\\
    &\equiv K\Omega_{\mu\nu} \cap \Z^2
\end{align}
where we define a line $\Omega_{\mu\nu}\subset \R^2$ as $\Omega_{\mu\nu} = \ker(M_{\mu}\vdashed M_{\nu})^T$ for convenience.\footnote{The case $\dim\Omega_{\mu\nu}=0$ is impossible under the assumption that bulk representatives of the confined sectors exist.}

The cylinder confinate is determined by two sets of boundary conditions (\ref{eq:em_bc}) corresponding to the top ($M_{\mu}$) and bottom ($M_{\nu}$) boundaries. So the cylinder confinate is 
\begin{align}
    \label{eq:confinate1}
    N_{\mu} \cap N_{\nu} &= \{\mb{n}\in\Z^2|\ M_{\mu}^T AA^T \mb{n}=0,\ M_{\nu}^T AA^T \mb{n}=0 \} \nonumber\\
    &=(AA^T)^{-1} \ker(M_{\mu}^T) \cap (AA^T)^{-1} \ker(M_{\nu}^T) \cap \Z^2 \nonumber\\
    &=(AA^T)^{-1} \left( \ker(M_{\mu}^T) \cap \ker(M_{\nu}^T) \right) \cap \Z^2 \nonumber\\
    &= (AA^T)^{-1} \ker(M_{\mu}\vdashed M_{\nu})^T \cap \Z^2 \nonumber\\
    &= (AA^T)^{-1}\Omega_{\mu\nu} \cap \Z^2
\end{align}

On the other hand, by definition the cylinder confinate is composed of the anyons that cannot escape from either boundary. In this sense it is generated by the orthogonal complement of all possible fusion results of the condensed anyons, i.e., $(\im(M_{\mu})\cup \im(M_{\nu}))^{\perp}$. There's a general relation between the kernel and image of a linear operator,  $\ker\mP^{T}=(\im\mP)^{\perp}$, from which the above expression can be simplified:
\begin{align}
    \label{eq:confinate2}
    N_{\mu} \cap N_{\nu} &=(\im(M_{\mu})\cup \im(M_{\nu}))^{\perp} \cap \Z^2 \nonumber\\
    &= \im(M_{\mu} \vdashed M_{\nu})^{\perp}\cap \Z^2 \nonumber\\
    &= \ker(M_{\mu}\vdashed M_{\nu})^T \cap \Z^2 \nonumber\\
    &= \Omega_{\mu\nu} \cap \Z^2
\end{align}

Comparing (\ref{eq:confinate1}) with (\ref{eq:confinate2}) we find that the line $\Omega_{\mu\nu}$ is an invariant subspace of $AA^T$:
\begin{equation}
    \label{eq:inv_subspace}
    AA^T \Omega_{\mu\nu} = \Omega_{\mu\nu}
\end{equation}

The self and mutual null conditions of the condensate impose a constraint on the confinfed subspace $\Omega_{\mu\nu}$:
\begin{equation}
    \label{eq:confinate_orthogonal}
    \Omega_{\mu\nu}^T K\Omega_{\mu\nu} =(K\Omega_{\mu\nu})^T K^{-1} (K\Omega_{\mu\nu})= 0
\end{equation}
From (\ref{eq:diagonalize},\ref{eq:inv_subspace} and \ref{eq:confinate_orthogonal}) we can also derive 
\begin{equation}
    \label{eq:condensate_orthogonal}
    \Omega_{\mu\nu}^T K^{-1} \Omega_{\mu\nu} =0
\end{equation}
It is easily observed here that the ``confined direction'' $\Omega_{\mu\nu}$ indeed satisfies the equation for ``condensed direction'' (\ref{eq:condensate_orthogonal}). For a fixed bulk topological order, the condensate $M$ and confinate $N$ for some boundary(given by boundary condensate $M$) become the confinate and condensate respectively for another boundary(given by boundary condensate $N$), and \textit{vice versa}. The roles played by the condensate and the confinate can be swapped. Given an unordered condensate-confinate pair ($M$, $N$), it's impossible to distinguish which is which without other information. We call this duality the \textit{condensed-confined duality}. It's a duality between the condensate and the confinate, or equivalently, between direction $K\Omega_{\mu\nu}$ and $\Omega_{\mu\nu}$. 

According to this duality, the relations appeared above(\ref{eq:inv_subspace},\ref{eq:confinate_orthogonal},\ref{eq:condensate_orthogonal}) still hold under the swapping $\Omega_{\mu\nu} \leftrightarrow K\Omega_{\mu\nu}$. In particular, $AA^T K\Omega_{\mu\nu} = K\Omega_{\mu\nu}$ and therefore $K^2\Omega_{\mu\nu}=\Omega_{\mu\nu}$. 

\begin{figure}
    \centering
    \includegraphics[width=0.3\textwidth]{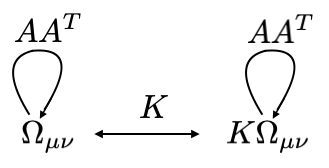}
    \label{fig:inv_subspace}
    \caption{$K\Omega_{\mu\nu}$ and $\Omega_{\mu\nu}$ are invariant subspaces of $AA^T$}
\end{figure}

The condensed-confined duality also appears in the non-Abelian $D(S_3)$ example, we refer the readers to appendix \ref{app:condensed_confined_DS3} for a detailed discussion.

Now we turn to several important 1D lattices appearing in the calculation of open string characters. We define
\begin{align}
    \Gamma:=\Omega_{\mu\nu} \cap K\Z^2,\qquad \tilde{\Gamma}:=K\Omega_{\mu\nu}\cap K\Z^2,
\end{align}
their \textit{dual lattices} are respectively given by
\begin{align}
    \Gamma^{\star}=\Omega_{\mu\nu} \cap K^{-1}\Z^2,\qquad \tilde{\Gamma}^{\star}=K\Omega_{\mu\nu}\cap K^{-1}\Z^2,
\end{align}
and satisfying
\begin{align}
    K\Gamma^{\star}=K\Omega_{\mu\nu} \cap \Z^2,\qquad K\tilde{\Gamma}^{\star}=\Omega_{\mu\nu}\cap \Z^2,
\end{align}

\begin{figure}[h]
    \centering
    \includegraphics[width=0.7\textwidth]{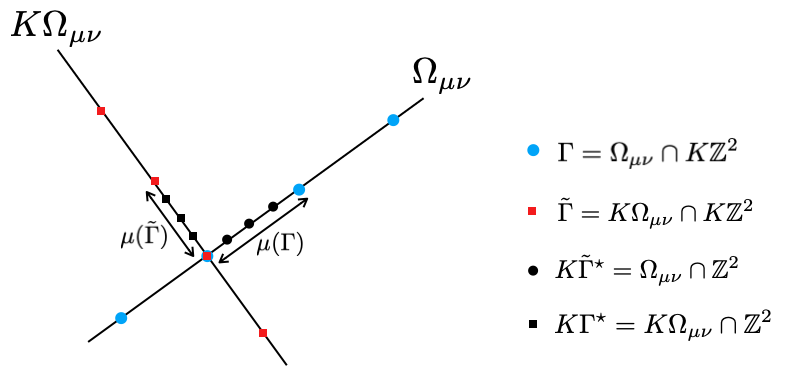}
    \label{fig:sublattice}
    \caption{1D sublattices of the charge lattice $\Z^2$}
\end{figure}

we have (see appendix \ref{app:poisson_resummation})
\begin{align}
    \label{eq:char_trans_1}
    \sum_{ \mb{m}\in\mb{x}+\Gamma }\exp(-\pi \mb{m}^T AA^T \mb{m}) 
    &=\frac{1}{\mu(\Gamma) \sqrt{p(A)}} \sum_{\mb{m}\in\Gamma^{\star}} \exp(-\pi \mb{m}^T (AA^T)^{-1} \mb{m} +2\pi i \mb{x}^T\mb{m} ) \nonumber\\
    &=\frac{1}{\mu(\Gamma) \sqrt{p(A)}} \sum_{\mb{n}\in K \Gamma^{\star}} \exp(-\pi \mb{n}^T K^{-1} (AA^T)^{-1} K^{-1} \mb{n} +2\pi i \mb{x}^T K^{-1} \mb{n} ) \nonumber\\
    &=\frac{1}{\mu(\Gamma) \sqrt{p(A)}} \sum_{c\in\mC} \sum_{\mb{n}\in\mb{c}+\tilde{\Gamma}} \exp(-\pi \mb{n}^TAA^T\mb{n} +2\pi i \mb{x}^T K^{-1}\mb{c}) \nonumber\\
    &= \frac{1}{\mu(\Gamma) \sqrt{p(A)}} \sum_{\mb{c}\in\mC} \exp(2\pi i \mb{x}^T K^{-1}\mb{c}) \sum_{\mb{n}\in\mb{c}+\tilde{\Gamma}} \exp(-\pi \mb{n}^TAA^T\mb{n})
\end{align}

where $\mC=M_{\mu}\cap M_{\nu}$, and  $p(A)$ is the scaling factor(eigenvalue) of the invariant subspace $\Omega_{\mu\nu}$:
\begin{equation}
     \forall \mb{m} \in \Omega_{\mu\nu},\quad AA^T \mb{m} = p(A) \mb{m},
\end{equation}

according to the condensed-confined duality, this factor $p(A)$ is also the scaling factor of the invariant subspace $K\Omega_{\mu\nu}$:
\begin{align}
\forall \mb{n} \in K\Omega_{\mu\nu},\quad AA^T \mb{n} = p(A) \mb{n},
\end{align}
    and the lattice spacing of $\Gamma$ is equal to the lattice spacing of $\tilde{\Gamma}$:
    \begin{equation}
        \mu(\Gamma)=\mu(\tilde{\Gamma})
    \end{equation}
so we also have
\begin{align}
    \label{eq:char_trans_2}
\sum_{\mb{n}\in\mb{c}+\tilde{\Gamma}} \exp(-\pi \mb{n}^TAA^T\mb{n})&=\frac{1}{\mu(\tilde{\Gamma})\sqrt{p(A)}}\sum_{\mb{n}\in\tilde{\Gamma}^{\star}} \exp(-\pi \mb{n}^T (AA^T)^{-1} \mb{n} +2\pi i \mb{c}^T\mb{n} ) \nonumber\\
    &=\frac{1}{\mu(\tilde\Gamma) \sqrt{p(A)}} \sum_{\mb{m}\in K \tilde{\Gamma}^{\star}} \exp(-\pi \mb{m}^T K^{-1} (AA^T)^{-1} K^{-1} \mb{m} +2\pi i \mb{c}^T K^{-1} \mb{m} ) \nonumber\\
    &=\frac{1}{\mu(\tilde\Gamma) \sqrt{p(A)}} \sum_{\mb{x}\in\mX} \sum_{\mb{m}\in\mb{x}+{\Gamma}} \exp(-\pi \mb{m}^TAA^T\mb{m} +2\pi i \mb{c}^T K^{-1}\mb{x}) \nonumber\\
    &= \frac{1}{\mu(\tilde\Gamma) \sqrt{p(A)}} \sum_{\mb{x}\in\mX} \exp(2\pi i \mb{c}^T K^{-1}\mb{x}) \sum_{\mb{m}\in\mb{x}+{\Gamma}} \exp(-\pi \mb{m}^TAA^T\mb{m})
\end{align}
where $\mX=N_{\mu}\cap N_{\nu}$

(\ref{eq:char_trans_1}) and (\ref{eq:char_trans_2}) give rise to the following transformation rules of the open/closed string characters:
\begin{align}
    \label{eq:char_trans_rules}
    \chi_{\mb{x}}^{(open)}=\sum_{\mb{c}\in\mC} \sigma_{\mb{x}\mb{c}} \chi_{\mb{c}}^{(closed)}, \quad \chi_{\mb{c}}^{(closed)}=\sum_{\mb{x}\in\mX } \sigma_{\mb{x}\mb{c}} \chi_{\mb{x}}^{(open)}
\end{align}
where
\begin{equation}
\sigma_{\mb{x}\mb{c}}=\frac{1}{\mu(\tilde\Gamma) \sqrt{p(A)}}\exp(2\pi i \mb{c}^T K^{-1}\mb{x}).
\end{equation}

This non-degenerate square matrix $\sigma$ is unitary (from (\ref{eq:char_trans_rules})) and satisfies
\begin{equation}
    \frac{\sigma_{\mb{x}\mb{c}}}{\sigma_{\mb{0}\mb{c}}}=\frac{S_{\mb{x}\mb{c}}}{S_{\mb{0}\mb{c}}}.
\end{equation}
So it's identified with $\gamma$ matrix \footnote{The $\gamma$ matrix for Abelian Chern-Simons theory is explicitly given in \cite{shen2019defect}, from which this relation is easily proved.}:
\begin{equation}
    \sigma_{\mb{x}\mb{c}}=\gamma_{\mb{x}\mb{c}}
\end{equation}

This result guarantees that the topological entanglement entropy where the entanglement cut is cutting across the gapped boundaries characterized by (\ref{eq:bc1}, \ref{eq:bc2}) is given again by (\ref{eq:topEE}).

\subsection{A note on the unitarity of $\gamma$ and more general gapped boundaries}\label{sec:generalizedEM}

In the previous discussions, we demonstrated the explicit construction of boundaries corresponding to different anyon condensation, and how the quantization of the edge modes at the entanglement cut leads to a reduced density matrix whose trace produces the twisted characters. The modular transformation of these twisted characters are determined by a set of half-linking matrix $\gamma$. If one assumes that $\gamma$ is unitary, then it is uniquely determined. In the previous sub-sections, we have presented explicit computations that recover a set of unitary half-linking matrix. We note that there are several choices we have made in our computation, that on hind-sight was responsible for the unitarity of the $\gamma$ matrix. Other choices could alter the overall normalization of the gamma matrices, thus shifting the topological entanglement by some non-universal factors. 

Specifically, the choice that we have made in the previous calculation is the value of the parameter $r$ that features in the effective action reviewed in equation (\ref{kmatrixZN}).
We have chosen $r=1$ in our computation and we note that this is a symmetric point that preserves the symmetry between $\phi_1$ and $\phi_2$ -- which is an electric-magnetic symmetry. While $r$ is canceled out in the computation of the topological entanglement entropy in the absence of boundaries, here it would change the overall normalization of the $\gamma$ matrix and shift the topological entanglement by 
\be
\gamma_{xc} \to \frac{\gamma_{xc}}{\sqrt{\tilde{r}}} , \qquad \Delta S_{EE} = - N \ln \tilde{r},
\ee
where $N$ is the number of entanglement cuts, and $\tilde{r}$ is the least positive number which makes $r\tilde{r}$ a perfect square.

     As a generalization of \ref{sec:EM}, we consider a cylinder with the bulk given by $\Z_{pq}$ gauge theories and the top and bottom boundaries characterized by subgroups $\Z_p$ and $\Z_q$ respectively, where $p$ and $q$ are relatively prime and both are greater than $2$ \cite{beigi_quantum_2011}.  These subgroups specify the subset of magnetic anyons in a Lagrangian algebra, which already uniquely specifies the condensates.
    The top boundary corresponds to the Lagrangian subgroup 
    $L_p=\left\langle
    \cvectorTwo{p}{0},
    \cvectorTwo{0}{q}\right\rangle$,
    while the bottom boundary corresponds to 
    $L_q=\left\langle
    \cvectorTwo{q}{0},
    \cvectorTwo{0}{p}\right\rangle$.
    This is a direct generalization of electric/magnetic boundary conditions considered above. Note that the condensed set $L_p$ (or $L_q$) does not satisfy the mutual null condition-- a condition necessary for defining a topological boundary condition for the Chern-Simons theory \cite{kapustin_topological_2011}.
   To cure the problem,  we extend the $2\times 2$ $K$ matrix to a $4\times 4$ symmetric integral matrix $\tilde{K}$ by adding 2 one-dimensional edge channels to the boundary \cite{levin_protected_2013,Barkeshli:2013yta}.
    \begin{eqnarray}
        \tilde{K}= K\oplus T=\left(
        \begin{array}{cccc}
            0 & pq & 0 & 0 \\
            pq & 0 & 0 & 0 \\
            0 & 0 & 0 & 1 \\
            0 & 0 & 1 & 0
        \end{array}    
        \right)
    \end{eqnarray}
    Adding $T=\left(\begin{array}{cc}
        0 & 1 \\
        1 & 0
    \end{array}\right)$ does not introduce any new quasiparticles, so $K$ and $\tilde{K}$ describe the same topological order.
    The Lagrangian subgroups 
    \begin{eqnarray}
        L_p=\left\langle \cvectorFour{p}{0}{0}{1} , \cvectorFour{0}{q}{-1}{0} \right\rangle, \quad
        L_q=\left\langle \cvectorFour{q}{0}{1}{0} , \cvectorFour{0}{p}{0}{-1} \right\rangle
    \end{eqnarray}
    are now generated by 4-dimensional charge vectors, satisfying the mutual null condition. For simplicity, we focus on the left side of one particular entanglement cut(superscript $l_i$) and supress this superscript hereafter. The boundary condition (\ref{BC}) becomes
    \begin{eqnarray}
        \label{eq:ZmZnBC}
        \begin{cases}
        (p\partial_t\phi_1+\partial_t\phi_4)|_{x=\pi}=0 \\
        (q\partial_t\phi_2-\partial_t\phi_3)|_{x=\pi}=0 \\
        (q\partial_t\phi_1+\partial_t\phi_3)|_{x=0}=0 \\
        (p\partial_t\phi_2-\partial_t\phi_4)|_{x=0}=0 
        \end{cases}
    \end{eqnarray}
    We now introduce the $4$ left/right moving modes according to decomposition $\tilde{K}=K\oplus T$
    \begin{eqnarray}
        &\phi_{1}=\sqrt{\frac{r^K}{2pq}}(\phi_L^K+ \phi_R^K),\quad
        &\phi_{2}=\frac{1}{\sqrt{2pqr^K}}(\phi_L^K- \phi_R^K), \nonumber \\
        &\phi_{3}=\sqrt{\frac{r^T}{2}}(\phi_L^T+ \phi_R^T), \quad
        &\phi_{4}=\frac{1}{\sqrt{2r^T}}(\phi_L^T- \phi_R^T).
    \end{eqnarray}
    To recover the conformal boundary condition at physical boundaries $x=0$ and $x=\pi$, we demand that (\ref{eq:ZmZnBC}) relates left-moving modes only to right-moving modes. This can only be achieved by tuning parameters to $r^K=1$, $r^T=\frac{q}{p}$. Then the boundary condition (\ref{eq:ZmZnBC}) reduces to conformal boundary conditions: 
    \begin{eqnarray}
        \label{eq:ZnZmCBC}
        \begin{cases}
    (\partial_t\phi_L^{K}-\partial_t\phi_R^{T})|_{x=\pi}=0 \\
    (\partial_t\phi_R^{K}+\partial_t\phi_L^{T})|_{x=\pi}=0 \\
    (\partial_t\phi_L^{K}+\partial_t\phi_R^{T})|_{x=0}=0\\
    (\partial_t\phi_R^{K}+\partial_t\phi_L^{T})|_{x=0}=0.
        \end{cases}
    \end{eqnarray}
    This conformal boundary condition admits only one solution out of all possible values of the zero modes, namely $P_1=a,P_2=-a,P_3=-qa,P_4=-pa$, or equivalently $P_L^K=P_R^T=0, P_R^K=-P_L^T=\sqrt{2pq}a$ where $a$ is an arbitrary integer.
    The mode expansion (\ref{eq:ModeExpansion}) together with the conformal boundary condition ({\ref{eq:ZnZmCBC}}) put the following constraint on the excitations:
    \begin{eqnarray}
        \label{eq:ZmZn_excitations}
        \left(
        \begin{array}{cccc}
            1 & 0 & 0 & -\lambda\\
            0 & \lambda & 1 & 0\\
            1 & 0 & 0 & 1\\
            0 & 1 & 1 & 0
        \end{array}  
        \right)
        \left(
        \begin{array}{c}
            a_n \\ b_n \\ c_n \\ d_n
        \end{array}    
        \right)=
        \left(
        \begin{array}{c}
            0 \\ 0 \\ 0 \\ 0
        \end{array}    
        \right)
    \end{eqnarray}
    where $\lambda=e^{2\pi in}$ and $a_n,b_n,c_n,d_n$ are excitations of $\phi_L^K,\phi_R^K,\phi_L^T,\phi_R^T$ respectively.
    The determinant of coefficients must vanish in order to have non-trivial excitations, so $(\lambda+1)(\lambda-1)=0$, or $n\in\mathbb{Z}\cup\{\frac{1}{2}+\mathbb{Z}\}$. (Compare with \textbf{e+m} case where $n$ is half-integer, and with \textbf{e+e}/\textbf{m+m} case where $n$ is integer.) 

    Following a similar procedure presented in \ref{sec:EM} we calculate the entanglement entropy between the strip regions $R$ and $\bar{R}$:
    \begin{equation}
        S=2(\frac{\pi l}{12\epsilon}-\log\sqrt{2} -\log\sqrt{2pq}).
    \end{equation}
   The extension of $K$ matrix and the precise choice of boundary conditions we have chosen in (\ref{eq:ZnZmCBC}) introduces a trapped Majorana mode, explaining the term $-2\log\sqrt{2}$. 
   We see that $r^T=\frac{q}{p}$ is a deviation from the symmetric point, leading to a shift $-\log\sqrt{pq}$ in the entanglement entropy. In other words, the half-linking matrix obtained here has a different normalization compared to the unitary one defined in \cite{shen2019defect}.

   Finally, one notices that there is one extra $-2\log \sqrt{2}$ attached alongside $-2\log{\sqrt{pq}}$. Physically, this had followed from the fact that we added an extra layer of "topologically trivial" material touching the $\mathbb{Z}_{pq}$ through a topological interface described by our boundary conditions (\ref{eq:ZmZnBC}). The interface induces topological symmetry enhancement in the trivial material, so that various trivial sectors in the "trivial material" becomes distinguishable through their connection with the non-trivial anyons. We have discussed this phenomenon already in \cite{lou2019ishibashi}. The topological entanglement gets contribution from both the $\mathbb{Z}_{pq}$ and $\mathbb{Z}_{1}$
   layers, leading to a factor of 2 in front of $pq$.      

    \section{Conclusion}
    
  In this paper, we demonstrate that the topological entanglement entropy is controlled by the ``half-linking'' number $\gamma_{xc}$ when the entanglement cut touches the gapped boundaries which are characterized by anyon condensation. We note that when the two gapped boundaries at the end of the entanglement cut correspond to two different anyon condensates, there could potentially be extra contribution to the topological entanglement  -- as illustrated by the case where the two ends of the entanglement cut  touch the electric and magnetic condensates. There is a non-trivial Majorana zero mode that contributes to a factor of $\sqrt{2}$. 
    
    One could consider more generic entanglement cuts that cut through gapped interfaces rather than boundaries. However, they could be understood in terms of gapped boundaries using the folding trick. 
   
Our computation was based on an ``open-string'' quantization at the entanglement cut. We supplement the open-string picture with a ``closed-string'' picture, by constructing a set of Cardy states suitable also for non-diagonal RCFT's. They are constructed using the half-linking numbers and reproduce the results based on the ``open-string'' computation. 
  We also prove, at least in the case of  Abelian Chern-Simons theories describable by 2-2 K-matrices, that there is a generalized notion of electric-magnetic duality. Namely there is a 1-1 correspondence between condensed and confined anyons and their charge vectors are related by a linear transformation. There is an analogous notion in non-Abelian theories with examples discussed in the appendix, although we could only hope for a precise proof in the future. 
  
  We note however, that the normalization of the half-linking matrix can be altered by some subtle change in the edge theory. We show that there is a class of choices which naturally preserve the unitarity of the half-linking matrix. This choice preserves the symmetry between electric and magnetic charges. 
  We also look into other choices of boundaries in which the half-linking matrix has a different overall normalization which departs from the unitary point. They lead to shifts in the normalization which enters into the topological entanglement. 
    
  One interesting direction we are currently pursuing is to generalize our work further to cases where the boundary remains gapless. Given the holographic/bulk-boundary correspondence properties found in topological orders which shares many similarities with the AdS/CFT correspondence, the computation should shed some further insight on the Ryu-Takayanagi formula.

    \appendix
    
    \section{Setting the notations of Abelian Chern-Simons theories} \label{CSreview}
    
    The class of Abelian Chern-Simons theories that we are going to consider in the following takes the following form: 
    \begin{equation}
    S_{CS}=\frac{1}{4\pi}\int_M K^{IJ}A_I\wedge F_J, \qquad F_J = dA_J.
    \end{equation}
    Where $M$ is a 3d manifold. Here $K_{IJ}$ is a symmetric integral matrix and $I=1,...,N$. 
    Quantization would involve gauge fixing (such as taking the temporal gauge $A^I_t=0$) and solving for the constraints following from the gauge choice. A review of its detailed procedure can be found for example in \cite{wang_boundary_2015}. Upon gauge fixing, the action becomes a total derivative. In the temporal gauge for example, the constraint equation would amount to the flat condition 
    \be F_{I\,\,xy} = 0. \ee 
    Setting 
    \be A_{I\,\,x,y} =\partial_{x,y} \phi_I \ee 
    for some scalar function $\phi$ and substituting these expressions into the bulk action, we recover a total derivative term. 
    For $M$ an open manifold with a 2d boundary $\partial M$, the total derivative gives rise to the following boundary action 
    \begin{equation} \label{CSaction}
    S_{\partial M}=\frac{{1}}{{4\pi}}\int_{\partial M} dtdx\,(K^{IJ}\partial_t\phi_I\partial_x\phi_J-V^{IJ}\partial_x\phi_I\partial_x\phi_J),
    \end{equation}
    There is an extra term involving an integral symmetric matrix $V^{IJ}$  of rank $m$. As discussed in \cite{wen_topological_1995}, it is not determined by the bulk CS action. They can be viewed as physical parameters that depend on the actual material supporting these gapless edge modes.  
    We note that $m$ being even is a necessary (although not sufficient) condition for the edge modes to be ``gappable'' by relevant perturbation. 
    The boundary action can be quantized canonically.
    This gives, at constant time $t$,
    \be
    [\phi_I(x),\Pi^J(y)] = i \delta^J_I\delta(x-y), \qquad \Pi^I(x)=\frac{{1}}{{2\pi}}K^{IJ}\partial_x\phi_J.
    \ee 
    Assuming that $x$ is compact and that $x \sim x+ l$ i.e. the boundary at constant time $t$ is a ring of length $l$. The mode expansion of $\Phi_I$ at $t=0$ is given by
    \begin{equation}
    \label{eq:ModeExpansion}
    \phi_I(x)=\phi_{0I}+K_{IJ}^{-1}P^{J}\frac{{2\pi}}{{l}}x+i\sum_{n\neq 0}\frac{{1}}{{n}}a_{I,n}e^{-inx\frac{{2\pi}}{{l}}}
    \end{equation}
    These modes therefore satisfy
    \be
    [\alpha_{I,n},\alpha_{J,m}]=nK_{IJ}^{-1}\delta_{n,-m}, 
    \ee 
    and for zero modes we have:
    $[\phi_{0I},P^{J}]=i\delta^{J}_I$.
    
    We will focus on the Chern-Simons equivalence of the $D(\Z_N)$ models in the following.  The corresponding $K$ matrix is given by
    \be \label{kmatrixZN}
    K= \begin{pmatrix}
    0&N\\N&0
    \end{pmatrix}.
    \ee
    The matrix has a pair of eigenvalues with opposite sign, signifying that it has exactly one pair of left and right moving modes, and as such,  is a non-chiral theory. 
    The scalars $\phi_I$ are related to the left and right moving fields by
    \begin{equation} \label{left_rightmodes}
    \phi_1=\sqrt{\frac{{r}}{{2N}}}(\phi_L+\phi_R), \qquad
    \phi_2=\sqrt{\frac{{1}}{{2Nr}}}(\phi_L-\phi_R)
    \end{equation}
    Here $r^2=V^{22}/V^{11}$.
    The left and right moving modes can also be expressed in a mode expansion:
    \be \label{expand_LR}
     \phi_{L(R)}(x)=\phi_{0L(R)}+P_{L(R)}\frac{{2\pi}}{{l}}x+i\sum_{n\neq 0}\frac{{1}}{{n}}\alpha_{L(R),n}e^{-inx\frac{{2\pi}}{{l}}}, \ee
     To avoid clutter, we will take $r=1$ in the following. We note that $r$ does not play any role in the topological entanglement of a single non-chiral phase. One can show that it is canceled out in the computation of the topological entanglement. It does play a non-trivial role in the discussion of generic interfaces between different $D(\Z_N)$ theories. We will re-introduce them where necessary.  We also note that when discussing topological entanglement in a chiral phase, one needs particular care in the choice of $r$. A detailed discussion will be taken up in the accompany paper. In that case, the entanglement cut crosses the physical interfaces, and extra care is needed. 
    Using the commutation relations of $\phi_I$, we recover
    \be \label{quantized_P}
    [\alpha_{L(R),n},\alpha_{L(R),m}]=n\delta_{n,-m}, \ee and
    \be \label{left_right_momenta}
     P_{L,R}=\frac{{1}}{{\sqrt{2N}}}(\pm P^{1}+P^{2}). \ee
    The $U(1)$'s gauge groups are taken to be compact. Therefore the scalars are also compact, satisfying
    \be 
    \phi_I \sim \phi_I + 2\pi.
    \ee
    The conjugate momenta to the zero modes therefore are quantized, satisfying
    \be \label{quantized}
    P^{I}\in\mathbb{Z}.
    \ee
    We note that these $P_{\phi_I}$ parametrizes a set of highest weight states. One can identify these highest weight states/operators with distinct anyons of the quantum double $D(\Z_N)$.  The identification with anyons is many-to-one: $P^{I}$ and $P^{I}+ N $ describe the same topological sector. One can take $P^{1} \mod N$ to parametrize the electric charge wrt to the $\Z_N$ gauge group in $D(\Z_N)$ models, and $P^{2}$ the magnetic charges. A detailed review can be found in \cite{wang_boundary_2015}. We only record the basic set of facts needed in the current paper.   
    The Hamiltonian is given by
    \begin{equation} \label{Hamiltonian}
    H=\frac{{1}}{{4\pi}}\int_0^l dx(\partial_x\phi_L \partial_x\phi_L+\partial_x\phi_R\partial_x\phi_R)=\frac{{P_L^2+P_R^2}}{{2}}+\sum_{n>0}(\alpha_{L,-n}\alpha_{L,n}+\alpha_{R,-n}\alpha_{R,n})-\frac{{1}}{{12}}
    \end{equation}

    \subsection{Review of gapped boundaries in $\Z_N$ theory} \label{app:revgapped}
    We briefly review gapped boundary and boundary conditions following \cite{lou2019ishibashi}.
    Recall that a gapped boundary is characterized by anyon condensation that takes the topological order A to the trivial phase.  The set of condensed anyons would form a so called Lagrangian algebra. This has been discussed in general in \cite{Bais:2002ny,Bais:2002pb, bais_condensate-induced_2009, Bais:2008xf,kapustin_topological_2011,
  barkeshli_classification_2013, Barkeshli:2013yta,
     Kong:2013aya, fuchs2013bicategories, levin_protected_2013, wang_boundary_2015, hung_ground_2015, hung_generalized_2015} and in the special case of Abelian CS theories, in \cite{kapustin_topological_2011,levin_protected_2013}.  In a $D(\Z_N)$ quantum double, all such Lagrangian algebras are known. We can take the set $L$ of condensed anyons as 
    \be  \label{eq:condensed_vecs}
    L = \{(P^{1}, P^{2})\},
    \ee
    where $(P^{1}, P^{2})$ is the pair of quantized quantum numbers (see equation (\ref{left_rightmodes}, \ref{left_right_momenta})) of the condensed sector.
    
    Among them there are two sets of gapped boundaries that are shared by all 
    $D(Z_N)$ theories and we will take them as examples for illustration purpose. These boundaries are called ``electric'' and ``magnetic'' boundaries respectively. Physically, the former correspond to the condensation of all electric charges and magnetic charges respectively i.e.
    \be \label{electricL}
    L_E = \{( N n + a, 0)\}, \qquad n \in \bZ, \qquad 0\leq a\leq N-1,
    \ee
    and similarly
    \be \label{magneticL}
    L_M = \{( 0, N m + b)\}, \qquad m \in \bZ, \qquad 0\leq b\leq N-1.
    \ee
    We note that these vectors that are collected into the condensed set $L$ are more selective than picking all charge vectors corresponding to the condensed topological sector. 
    In particular, they are ``self-null'' and ``mutually-null'' vectors satisfying 
    \be \label{null_condition}
    P_{l_i}^I K^{-1}_{IJ} P_{l_j}^J = 0, \qquad  \forall (P^1_{l_i}, P^2_{l_i}) \in L_{c},
    \ee
    where $L_c$ denotes a generic collection of vectors of condensates in a Lagrangian algebra.  This has been discussed at length in \cite{levin_protected_2013,wang_boundary_2015}, particularly how they are related to existence of corresponding relevant operators that could gap these edge modes. 
    
    Alternatively, one can think of these Lagrangian algebra as characterizing conformal boundary conditions \cite{cardy_boundary_1989,cardy_boundary_2004}. We note that the boundary theory has a set of  $U(1)$ global symmetries extended to a $U(1)$ Kac-Moody algebra.  The conserved currents are given by
    \be
    J^I_{x} = \frac{K^{IJ}}{2\pi} \partial_x \phi_J. 
    \ee
    
 This implies that the zero mode of the current is given by
 \be \label{J0}
 J^I_{x,0} \equiv \int_0^l dx \, J^I_{x} = P^I.
 \ee   
    The Lagrangian algebra defines a boundary condition, or alternatively a boundary state $|\psi\rangle\rangle$ that preserves the following symmetries
    \begin{equation}\label{BC}
    P_{l_i}^I K^{-1}_{IJ} J^J_x |\psi\rangle\rangle=0.
    \end{equation}
    Using (\ref{J0}), this implies that the boundary condition is allowing the state to carry non-trivial expectation values of $P_{I_i}$ simultaneously if they are mutually null, as described in (\ref{null_condition}).  
    Indeed we only need a minimal set of vectors $(P^1_{l_i}, P^2_{l_i})$ that are linearly independent to generate the entire $L_c$.
    In the case of the electric boundary, we need only the null vector $(1,0)$. i.e.
    \be \label{celectric}
    K^{-1}_{12} J^2_x  |\psi\rangle\rangle_E \equiv \frac{1}{2\pi}\partial_x \phi_1 |\psi\rangle\rangle_E =0.
    \ee
    Similarly a magnetic boundary would amount to taking the condensate vector  $(0,1)$, leading to 
    \be \label{cmagnetic}
    \frac{1}{2\pi}\partial_x \phi_2 |\psi\rangle\rangle_M =0.
    \ee
    Now in terms of the right and left moving fields, the above conditions on the boundary state can be re-written as
    \be \label{Jbcstate}
    (J_L  \pm J_R)|\psi\rangle\rangle_{E/M}=0, \qquad  J_{L,R} = \frac{1}{2\pi} \partial_x \phi_{L,R},
    \ee
    where $\phi_{L,R}$ are related to $\phi_{1,2}$ by (\ref{left_rightmodes}). We immediately note that the above equations implies that the states $|\psi\rangle\rangle_{E/M}$ 
    are indeed conformal boundary states satisfying the conformal boundary condition,
    \be
    (L_n - \bar L_{-n}) |\psi\rangle\rangle_{E/M} = 0,
    \ee
    where $L_n$ are the Virasoro generators of the left-moving modes and $\bar L_m$ the corresponding generators of the right-moving modes. 
    This follows from the fact that the stress tensor can be expressed as 
    \be
    T = \pi J_L J_L, \qquad \bar T =\pi J_R J_R
    \ee 
    by the Sugawara construction. We note that the Hamiltonian $H$ in (\ref{Hamiltonian}) is indeed given by
    \be
    H = L_0 + \bar L_0 - \frac{c}{12}, \qquad  \textrm{where} \,\, c=1.
    \ee
    In terms of the mode expansion, 
    \be
    (\alpha_{{L},n}\pm \alpha_{{R},-n})|\psi\rangle\rangle_{E/M}=0
    \ee
    The corresponding boundary Ishibashi state has the following form:
    \begin{equation}
    |\psi\rangle\rangle_{E/M}= \exp(- \frac{2\pi\epsilon}{l} H) \exp(\mp \sum^{\infty}_{n=1}\frac{{1}}{{n}}\alpha_{{L},-n}\alpha_{{R},-n}))|P_L,P_R\rangle\rangle_{E/M},
    \end{equation}
    where 
    \be
    P_L = \mp P_R,
    \ee
    for electric and magnetic boundaries respectively. 
    The boundary state is not normalizable, and so $\exp(- \frac{2\pi\epsilon}{l} H)$ serves as a regularization, with $\epsilon$ infinitesimal. The parameter $l$ is the length of the circle. 
    The norm of this state is then given by
    \be
    \langle \langle \psi | \psi \rangle \rangle = \frac{q^\frac{P_L^2}{2}}{\eta(q)}, \qquad q=e^{\frac{-8\pi\epsilon}{l}}
    \ee
    
    This has been discussed for example in \cite{fliss_interface_2017}, although we would like to make the connection to anyon condensation more transparent in the current discussion.


    \section{Some useful details of the $D(S_3)$ model} \label{S3model_review}
    We would like to review here some basic data of the $D(S_3)$ model. 
    The anyons are labeled by $(C,\rho_{\alpha_C})$, where $C$ is a conjugacy class of the group $G=S_3$, and $\alpha_C$ an irrep of the centralizer of $C$. A summary of all the anyons are listed below. 
    
        \begin{tabular}{c|ccc|cc|ccc}
            \hline\hline
             & $A$ & $B$ & $C$ & $D$ & $E$ & $F$ & $G$ & $H$ \\
            \hline
            conjugacy class $W$ & \multicolumn{3}{|c|}{$\{e\}$} & \multicolumn{2}{c|}{$\{y,xy,x^2y\}$} & \multicolumn{3}{c}{$\{x,x^2\}$} \\
            \hline
            centralizer $\cong$ & \multicolumn{3}{|c|}{$S_3$} & \multicolumn{2}{c|}{$\Z_2$} & \multicolumn{3}{c}{$\Z_3$} \\
            \hline
            irrep $\rho$ of centralizer  & \boldmath$1$ & sign & \boldmath$\pi$ & \boldmath$1$ & \boldmath$-1$ & \boldmath$1$ & \boldmath$\omega$ & \boldmath$\omega^*$ \\
            \hline
            dim($\rho$) & 1 & 1 & 2 & 1 & 1 & 1 & 1 & 1 \\
            \hline
            quantum dimension $d=|W|\times$ dim$(\rho)$ & 1 & 1 & 2 & 3 & 3 & 2 & 2 & 2\\
            \hline
            twist $\theta$ & 1 & 1 & 1 & 1 & -1 & 1 & $e^{2\pi i/3}$ & $e^{-2\pi i/3}$\\
            \hline\hline
        \end{tabular} 
    
    Their fusion rules are given by
    
        \scalebox{0.7}{
        \begin{tabular}{c|ccc|cc|ccc}
            \hline\hline
            $\otimes$ & $A$ & $B$ & $C$ & $D$ & $E$ & $F$ & $G$ & $H$ \\ \hline
            $A$ & $A$ & $B$ & $C$ & $D$ & $E$ & $F$ & $G$ & $H$ \\ 
            $B$ & $B$ & $A$ & $C$ & $E$ & $D$ & $F$ & $G$ & $H$ \\ 
            $C$ & $C$ & $C$ & $A\oplus B\oplus C$ & $D\oplus E$ & $D\oplus E$ & $G\oplus H$ & $F\oplus H$ & $F\oplus G$ \\ \hline
            $D$ & $D$ & $E$ & $D\oplus E$ & $A\oplus C\oplus F\oplus G\oplus H$ & $B\oplus C\oplus F\oplus G\oplus H$ & $D\oplus E$ & $D\oplus E$ & $D\oplus E$ \\ 
            $E$ & $E$ & $D$ & $D\oplus E$ & $B\oplus C\oplus F\oplus G\oplus H$ & $A\oplus C\oplus F\oplus G\oplus H$ & $D\oplus E$ & $D\oplus E$ & $D\oplus E$ \\ \hline
            $F$ & $F$ & $F$ & $G\oplus H$ & $D\oplus E$ & $D\oplus E$ & $A\oplus B\oplus F$ & $C\oplus H$ & $C\oplus G$ \\ 
            $G$ & $G$ & $G$ & $F\oplus H$ & $D\oplus E$ & $D\oplus E$ & $C\oplus H$ & $A\oplus B\oplus G$ & $C\oplus F$ \\
            $H$ & $H$ & $H$ & $F\oplus G$ & $D\oplus E$ & $D\oplus E$ & $C\oplus G$ & $C\oplus F$ & $A\oplus B\oplus H$ \\
            \hline\hline
        \end{tabular}
        }

    The $S$-matrix is given by
    \begin{eqnarray}
        \label{eq:S3_S_matrix}
        S=\frac{1}{6}\left(
        \begin{array}{cccccccc}
           1 & 1 & 2 & 3 & 3 & 2 & 2 & 2 \\
           1 & 1 & 2 & -3 & -3 & 2 & 2 & 2 \\
           2 & 2 & 4 & 0 & 0 & -2 & -2 & -2 \\
           3 & -3 & 0 & 3 & -3 & 0 & 0 & 0 \\
           3 & -3 & 0 & -3 & 3 & 0 & 0 & 0 \\
           2 & 2 & -2 & 0 & 0 & 4 & -2 & -2 \\
           2 & 2 & -2 & 0 & 0 & -2 & -2 & 4 \\
           2 & 2 & -2 & 0 & 0 & -2 & 4 & -2 \\
        \end{array}    
        \right),
    \end{eqnarray}

    \section{Condensed-Confined duality of $D(S_3)$}\label{app:condensed_confined_DS3}
    There're $4$ distinct gapped boundaries for a bulk theory $D(S_3)$, labeled by the $4$ different subgroups of $S_3$. The condensed anyons corresponding to each subgroup are listed in Table \ref{tab:condense}.
    \begin{table}[h]
        \centering
    \begin{tabular}{c|c}
        \hline\hline
        subgroup $K$ & condensate \\
        \hline
        $\mb{1} $ & $A\oplus B\oplus 2C$ \\
        $\Z_2$ & $A\oplus C\oplus D$ \\
        $\Z_3$ & $A\oplus B\oplus 2F$ \\
        $S_3$ & $A\oplus D\oplus F$ \\
        \hline\hline
    \end{tabular}
    \caption{Condensed anyons corresponding to boundary subgroup $K$}
    \label{tab:condense}
    \end{table}

    The condensates are not independent. What is previously known is the $C\leftrightarrow F$ duality: a new condensate can be obtained by swapping $C$ and $F$ in a given condensate. We observe here that there's another relation among the condensates. For the gapped boundary $\mathcal{A}_4=A\oplus D\oplus F$, the confined sectors are labeled by $A$, $B$ and $C$, which are exactly the condensed anyons for the gapped boundary $\mathcal{A}_1=A\oplus B\oplus 2C$. The situation is more complicated in the inverse direction: for the gapped boundary $\mathcal{A}_1=A\oplus B\oplus 2C$, the confined sectors are labeled by $A$, $D$ and $F$ with appropriate idempotent splitting.\cite{cong_topological_2016}. 
    \begin{figure}[h]
        \centering
        \includegraphics[width=0.7\textwidth]{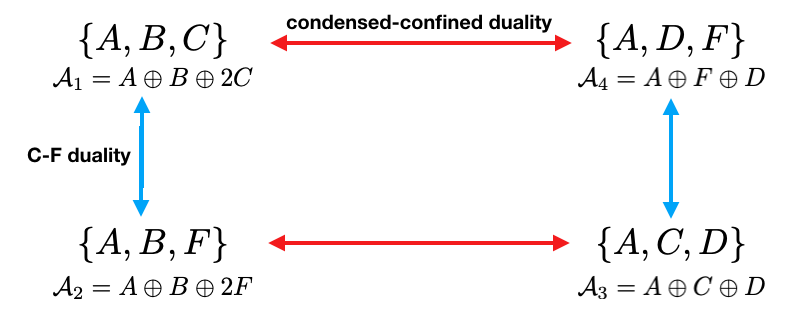}
        \label{fig:S3}
        \caption{The $4$ boundary condensates of $D(S_3)$ Dijkgraaf-Witten model, related by the $C\leftrightarrow F$ duality and the condensed-confined duality.}
    \end{figure}
    
    \section{$\eta$ and $\theta$ functions}
    \label{app:eta_theta_function}
    We list here the definitions and basic properties of Dedekind $\eta-$function and Jacobi $\theta-$function.
        \begin{eqnarray}
            \eta(\tau)=q^{\frac{1}{24}}\prod_{n=1}^{\infty}(1-q^n),
        \end{eqnarray}
        \begin{eqnarray}
            \theta_2(\tau)=\sum_{n\in\mathbb{Z}}^{}q^{\frac{1}{2}(n+\frac{1}{2})^2}&=&2\eta(\tau)q^{\frac{1}{12}}\prod_{r=1}^{\infty}(1+q^r)^2, \nonumber\\
            \theta_3(\tau)=\sum_{n\in\mathbb{Z}}^{}q^{\frac{n^2}{2}}&=&\eta(\tau)q^{-\frac{1}{24}}\prod_{r=0}^{\infty}(1+q^{r+\frac{1}{2}})^2, \nonumber\\
            \theta_4(\tau)=\sum_{n\in\mathbb{Z}}^{}(-1)^nq^{\frac{n^2}{2}}&=&\eta(\tau)q^{-\frac{1}{24}}\prod_{r=0}^{\infty}(1-q^{r+\frac{1}{2}})^2
        \end{eqnarray}
        where $\tau$ is the modular parameter and $q=e^{2\pi i \tau}$. 

        These functions are related by modular $T$ transformation ($\tau\to\tau+1$) and $S$ transformation ($\tau\to-\frac{1}{\tau}$):
        \begin{eqnarray}
            \eta(\tau+1)=e^{\frac{\pi i}{12}}\eta(\tau),& \qquad \eta(-\frac{1}{\tau})=\sqrt{-i\tau}\eta(\tau)\nonumber\\
            \theta_2(\tau+1)=e^{\frac{\pi i}{4}}\theta_2(\tau),&\qquad \theta_2(-\frac{1}{\tau})=\sqrt{-i\tau}\theta_4(\tau), \nonumber\\
            \theta_3(\tau+1)=\theta_4(\tau),&\qquad \theta_3(-\frac{1}{\tau})=\sqrt{-i\tau}\theta_3(\tau), \nonumber\\
            \theta_4(\tau+1)=\theta_3(\tau),&\qquad \theta_4(-\frac{1}{\tau})=\sqrt{-i\tau}\theta_2(\tau)
        \end{eqnarray}
        In the main text $\tilde{q}$ is defined as the $S$ transformation of corresponding $q$, for any complex number $X$
        \begin{equation}
            q=e^X\xrightarrow{S}\tilde{q}=e^{\frac{4\pi^2}{X}}
        \end{equation}

    \section{Poisson resummation}
        \label{app:poisson_resummation}
        The Poisson resummation formula is a beautiful relation between a function $f(x)$ and its Fourier transform $\hat{f}(y)=\int e^{2\pi i \langle x,y\rangle} f(x)dx$, stating that the following infinite sums are equal:
        \begin{equation}
            \label{eq:original_poisson}
            \sum_{n\in\Gamma} f(n) = \frac{1}{\mu(\Gamma)}\sum_{n\in\Gamma^{\star}} \hat{f}(n)
        \end{equation}
        where $\Gamma$ is an 1D lattice (a free $\Z$-module) and $\mu(\Gamma)$ its lattice spacing (measure of unit cell). $\Gamma^{\star}$ is the \textit{dual lattice} of $\Gamma$, defined as
        \begin{equation}
            \Gamma^{\star}=\{ n\in\R | \langle m,n \rangle\in\Z,\ \forall m\in\Gamma \}.
        \end{equation}
        In 1D the inner product $\langle \cdot,\cdot\rangle$ is reduced to multiplication in $\R$.
        In particular the following special case is the most frequently invoked in physics literature:
        \begin{equation}
        \sum _ { n \in \mathbb { Z } } \exp \left( - \pi a n ^ { 2 } + b n \right) = \frac { 1 } { \sqrt { a } } \sum _ { k \in \mathbb { Z } } \exp \left( - \frac { \pi } { a } \left( k + \frac { b } { 2 \pi i } \right) ^ { 2 } \right)
        \end{equation}
        and its inverse
        \begin{equation}
            \sum_{k\in\Z} \exp(-\pi a (k+b)^2)=\frac{1}{\sqrt{a}} \sum_{n\in\Z} \exp(-\frac{\pi}{a}n^2+2\pi i bn)
        \end{equation}
        The generalization of (\ref{eq:original_poisson}) to higher dimension is obvious.\cite{serre_cours_1994}
        
    \section{Half-linking $\gamma$ matrix} 
    \label{app:gamma_matrix}
    We briefly review $\gamma$ matrix following \cite{shen2019defect}.
    For a topological order on a cylinder with gapped boundaries $\mu$ and $\nu$, there are naturally two sets of basis to describe the ground state subspace, namely the Wilson loop basis $\ket{\wloop{x}}$ labeled by the confined sector $x$, and the Wilson line basis $\ket{\wline{c}}$ labeled by the condensed sector $c$ shared by the top and bottom physical boundaries. We label by $\mathcal{C}$ the set of shared condensed sectors.
    
    The $\gamma$ matrix is the transformation matrix between the two sets of basis. 
    \begin{equation}
        \label{def:gamma}
    \ket{{\wloop{x}}}=\sum_{c\in \mathcal{C}}\gamma_{xc}\ket{{\wline{c}}}
    \end{equation}
    

    The unitarity of $\gamma$ matrix comes from the orthonormality of the Wilson line basis and the Wilson loop basis:

    \begin{eqnarray*} \delta_{xy}&=&\braket{{\wloop{x}}}{{\wloop{y}}} \\ &=&
    \sum_{c,d}\bra{{\wline{d}}}\gamma_{dy}^{\dagger}\gamma_{xc}\ket{{\wline{c}}} \\ &=& \sum_{c,d}\gamma_{dy}^{\dagger}\gamma_{xc}\delta_{cd}=\sum_{c}\gamma_{cy}^{\dagger}\gamma_{xc}.
     \end{eqnarray*}

From the unitarity we can obtain the inverse transformation:
\[ \ket{{\wline{c}}}=\sum_{x}\gamma_{cx}^{\dagger}\ket{{\wloop{x}}}=\sum_{x}\overline{\gamma_{xc}}\ket{{\wloop{x}}} \]


    
     \subsection{$\mB=\mZ(\mC)$}
    If the bulk theory $\mB=\mZ(\mC)$ can be factorized $\mB=\mC\boxtimes\bar{\mC}$ where $\mC$ is also a modular tensor category with modular matrix $S^{\mC}$, then $\gamma_{xc}=S^{\mC}_{x,c}=\frac{S^{\mB}_{x\bar{0},c\bar{c}}}{\sqrt{S^{\mB}_{0\bar{0},c\bar{c}}}}$

    We can perform the folding/unfolding trick if $\mB$ can be factorized. Unfolding the cylinder to a torus, the doubled theory $\mB$ splits into $\mC$ and $\bar{\mC}$. The Wilson loop $\wloop{x}$ unfolds to $\wloop{x}$ and $\wloop{\bar{0}}$, while the Wilson line $\wline{c}$ unfolds to $\wline{c}$ and $\wline{\bar{c}}$.

    The $S^{\mB}$ for the doubled theory is the tensor product $S^{\mB}=S^{\mC}\otimes S^{\bar{\mC}}$, so \[ S^{\mB}_{i\bar{j},k\bar{l}}=S^{\mC}_{ik} S^{\bar{\mC}}_{\bar{j}\bar{l}}=S^{\mC}_{ik}S^{\mC}_{jl}.\] The last step comes from the fact $S^{\bar{\mC}}_{\bar{j}\bar{l}}=S^{\mC}_{jl}$. 
    Hence we have
    \begin{eqnarray}
    \frac{S^{\mB}_{x\bar{0},c\bar{c}}}{\sqrt{S^{\mB}_{0\bar{0},c\bar{c}}}}=S^{\mC}_{x,c}
    \end{eqnarray}

     The $\gamma$ matrix can be identified with the $S^{\mC}$ matrix in this case. Because the cylinder basis transformation $\ket{\wloop{x}}\leftrightarrow\ket{\wline{c}}$, if viewed in the unfolded picture, is exactly the basis transformation $\ket{\wloop{x}}\leftrightarrow\ket{\wloopVtc{c}}$ on a torus, which is dictated by $S^{\mC}$ matrix.

  
 \subsection{Abelian Chern Simons}
If the bulk theory is Abelian Chern Simons theory given by the $K$ matrix, and both boundaries of the cylinder are described by condensate $C$,  then $\gamma_{xc}=\frac{S_{xc}}{\sqrt{S_{0c}}}$.

\begin{equation}
\ket{\wloop{0}}=\frac{1}{\sqrt{D}}\sum_{c\in\mC}\ket{\wline{c}}
\end{equation}
where the normalization constant $\frac{1}{\sqrt{D}}$ comes from $|\mC|=D$.
Applying Wilson loop operators to this state
\begin{eqnarray}
\ket{\wloop{x}}=\widehat{\wloop{x}}\ket{\wloop{0}}&=&\frac{1}{\sqrt{D}}\sum_{c\in\mC}\widehat{\wloop{x}}\ket{\wline{c}}\nonumber\\
&=&\frac{1}{\sqrt{D}}\sum_{c\in\mC}\frac{S_{xc}}{S_{0c}}\ket{\wline{c}}\nonumber\\
&=&\frac{S_{xc}}{\sqrt{S_{0c}}}\ket{\wline{c}}
\end{eqnarray}
in the last line we've used $S_{0a}=\frac{d_a}{D}=\frac{1}{D}$ for abelian Chern Simons theory.


 \section{TQFT S matrix = CFT S matrix}
For Abelian Chern-Simons theory on a torus, the \text{TQFT} $S$ matrix from anyon braiding is identified with \text{CFT} $S$ matrix from modular transformation of characters.

If an Abelian Chern Simons theory supports gapped edge, then its bulk $K$ matrix must have total signature $0$, and hence congruent to $\matrixTwo{\one}{0}{0}{-\one}$. 
Suppose 
\begin{eqnarray}
\label{eq:K_cong}
A^{T}KA=\matrixTwo{\one}{0}{0}{-\one},
\end{eqnarray}
taking the determinant of both sides gives $\det A=\frac{1}{\sqrt{|\det{K}|}}=\frac{1}{D}$, where $D$ is the total quantum dimension. In abelian Chern Simons theory, each anyon has quantum dimension $1$, so $D^2=\text{\#(anyon types)}=|\det{K}|$.
The quantized conjugate momentum is:\footnote{For simplicity and without loss of generality, we will assume $K$ to be a $2\times 2$ matrix, the generalization is obvious.}
\begin{eqnarray}
\cvectorTwo{P_1}{P_2}=K^{-1}\mb{m},
\end{eqnarray}
where $\mb{m}= (P^1,P^2)^T$ is an integer vector. Transforming to the left/right-moving frame,
\begin{eqnarray}
\cvectorTwo{P_L}{P_R}=A^{-1}\cvectorTwo{P_1}{P_2}=(KA)^{-1}\mb{m}
\end{eqnarray}
Whenever we want to calculate the character of a CFT living on an entanglement cut, the pattern $\sum e^{-2\pi H}$ always appear. If we analyze the Hamiltonian closely, 
\begin{eqnarray}
    \label{eq:Hamiltonian}
H=\frac{{P_L^2+P_R^2}}{{2}}+\sum_{n>0}(\alpha_{L,-n}\alpha_{L,n}+\alpha_{R,-n}\alpha_{R,n})-\frac{{1}}{{12}}
\end{eqnarray}
the first part $\frac{{P_L^2+P_R^2}}{{2}}$ is responsible for Poisson resummation and yields Jacobi $\theta$ function, while the second part $\sum_{n>0}(\alpha_{L,-n}\alpha_{L,n}+\alpha_{R,-n}\alpha_{R,n})-\frac{{1}}{{12}}$ always gives Dedekind $\eta$ function. The summation in $\sum e^{-2\pi H}$ is performed over all primaries inside a superselection sector, which is reduced to summation over some integer lattice sites equivalent under $K$. Denote this lattice by $\Gamma$, it's a sublattice of $\Z^2$, and satisfies 
\begin{eqnarray}
 \forall \mb{x},\mb{y}\in\Gamma, \exists \mb{n}\in\Z^2 \text{ such that } \mb{x}-\mb{y}=K\mb{n}
\end{eqnarray}

The first part of the Hamiltonian(\ref{eq:Hamiltonian}) can be written as
\begin{eqnarray}
\frac{1}{2}(P_L^2+P_R^2)&=&\frac{1}{2}\rvectorTwo{P_L}{P_R}\matrixTwo{1}{0}{0}{1}\cvectorTwo{P_L}{P_R}\nonumber\\
&=&\frac{1}{2}\mb{m}^{T}((KA)^{-1})^{T}\matrixTwo{1}{0}{0}{1}(KA)^{-1}\mb{m}\nonumber\\
&=&\frac{1}{2}\mb{m}^{T}AA^T\mb{m}.
\end{eqnarray}

The 1D Poisson resummation formula 
\[
\sum_{n\in\Z} \exp(-\pi an^2+bn)=\frac{1}{\sqrt{a}} \sum_{k\in\Z}\exp(-\frac{\pi}{a}(k+\frac{b}{2\pi i})^2) \]

can be easily generalized to 2D:
\[
 \sum_{\mb{n}\in\Z^2} \exp(-\pi \mb{n}^T A \mb{n}+\mb{b}^T \mb{n}) = \frac{1}{\sqrt{\det A}}\sum_{\mb{k}\in\Z^2} \exp(-\pi (\mb{k}+\frac{\mb{b}}{2\pi i})^T A^{-1} (\mb{k}+\frac{\mb{b}}{2\pi i})) \]

Consider $\Gamma_{\mb{a}}=K\Z^2+\mb{a}$, let's focus on the Poisson resummation part of $\exp(-2\pi H)$
\begin{eqnarray}
&&\sum_{\mb{m}\in\Gamma_a}\exp(-\pi\mb{m}^T AA^T \mb{m})\nonumber\\
&=&\sum_{\mb{n}\in\Z^2}\exp(-\pi (K\mb{n}+\mb{a})^T AA^T (K\mb{n}+\mb{a}))\nonumber\\
&=& \sum_{\mb{n}\in\Z^2}\exp(-\pi\mb{n}^T A^{-T}A^{-1}\mb{n}-2\pi\mb{a}^T AA^T K\mb{n}-\pi \mb{a}^T AA^T\mb{a})\nonumber\\
&=&\frac{1}{\sqrt{\det(A^{-T}A^{-1})}}\sum_{\mb{k}\in\Z^2}\exp(-\pi(\mb{k}+iKAA^T \mb{a})^T AA^T (\mb{k}+iKAA^T \mb{a})-\pi\mb{a}^T AA^T\mb{a})\nonumber\\
&=&\det(A)\sum_{\mb{k}\in\Z^2}\exp(-\pi\mb{k}^T AA^T \mb{k}-2\pi i\mb{a}^TAA^TKAA^T\mb{k})
\end{eqnarray}
where in the second line we rewrite $\mb{m}=K\mb{n}+\mb{a}$ and transform the sum over $\mb{m}$ to the sum over $\mb{n}$. 
As demonstrated above, the next key step is to split the sum $\sum_{\mb{k}\in\Z^2}$ to a double sum. To this end, we rewrite $\mb{k}=K\mb{p}+\mb{b}$. In order to recover the $\Z^2$ lattice over which $\mb{k}$ is summed, $\mb{b}$ must run through all possible anyon types, namely $\sum_{\mb{k}\in\Z^2}=\sum_{\mb{p}\in\Z^2}\sum_{\mb{b}\in L}$. Furthermore $AA^TKAA^TK=\one$ by the definition of $A$.
\begin{eqnarray}
&&\sum_{\mb{m}\in\Gamma_a}\exp(-\pi\mb{m}^T AA^T \mb{m})\nonumber\\
&=&\det(A)\sum_{\mb{b}\in L} \exp(-2\pi i \mb{a}^TK^{-1}\mb{b}) \sum_{\mb{p}\in\Z^2}\exp(-\pi (K\mb{p}+\mb{b})^T AA^T(K\mb{p}+\mb{b}))\nonumber\\
&=&\sum_{\mb{b}\in L}\frac{1}{D}\exp(-2\pi i \mb{a}^TK^{-1}\mb{b})\sum_{\mb{m}\in\Gamma_{\mb{b}}}\exp(-\pi \mb{m}^T AA^T \mb{m})\nonumber\\
&=&\sum_{\mb{b}\in L}S_{\mb{a}\mb{b}} \sum_{\mb{m}\in\Gamma_{\mb{b}}}\exp(-\pi \mb{m}^T AA^T \mb{m}),
\end{eqnarray}
where $S_{\mb{a}\mb{b}}=\frac{1}{D}\exp(-2\pi i \mb{a}^TK^{-1}\mb{b})$ is the TQFT $S$ matrix defined as the Hopf link with the two cycles labeled by $a$ and $b$ respectively.
Restore the full character by completing the $\eta$ function part and the appropriate modular parameter we get 
\begin{equation}
\chi_{\mb{a}}(q)=\sum_{\mb{b}\in L}S_{\mb{a}\mb{b}}\chi_{\mb{b}}(\tilde{q})
\end{equation}
Thus we have shown explicitly that the modular transformation of characters is effected by the TQFT $S$ matrix in Abelian Chern-Simons theory.

    \section*{Acknowledgements}
    We thank Laurent Freidel, Gabriel Wong and Yang Zhou for helpful discussions. We would also like to thank Juven Wang for many past conversations on the subject.
    We are particularly grateful to Yidun Wan for a critical reading of our draft. 
    Part of this work is done during LYH and CS's visit to Perimeter Institute, as part of the Emmy-Noether Fellowship programme. 
    LYH acknowledges the support of Fudan University and the Thousands Young Talents Program. This work is supported by the NSFC grant number 11875111.

    \bibliography{ref}

\providecommand{\href}[2]{#2}\begingroup\raggedright\begin{thebibliography}{10}

\bibitem{Wang:2018edf}
J.~Wang, K.~Ohmori, P.~Putrov, Y.~Zheng, Z.~Wan, M.~Guo, H.~Lin, P.~Gao, and
  S.-T. Yau, ``{Tunneling Topological Vacua via Extended Operators: (Spin-)TQFT
  Spectra and Boundary Deconfinement in Various Dimensions},''
  \href{http://dx.doi.org/10.1093/ptep/pty051}{{\em PTEP} {\bfseries 2018}
  no.~5, (2018) 053A01},
\href{http://arxiv.org/abs/1801.05416}{{\ttfamily arXiv:1801.05416
  [cond-mat.str-el]}}.

\bibitem{Shi:2018krj}
B.~Shi and Y.-M. Lu, ``{Characterizing topological order by the information
  convex},'' \href{http://dx.doi.org/10.1103/PhysRevB.99.035112}{{\em Phys.
  Rev.} {\bfseries B99} no.~3, (2019) 035112},
\href{http://arxiv.org/abs/1801.01519}{{\ttfamily arXiv:1801.01519
  [cond-mat.str-el]}}.

\bibitem{chen_entanglement_2018}
C.~Chen, L.-Y. Hung, Y.~Li, and Y.~Wan, ``Entanglement {Entropy} of
  {Topological} {Orders} with {Boundaries},'' {\em arXiv:1804.05725 [cond-mat,
  physics:hep-th]} (Apr., 2018) . \url{http://arxiv.org/abs/1804.05725}. 00001
  arXiv: 1804.05725.

\bibitem{Shi:2018bfb}
B.~Shi, ``{Seeing topological entanglement through the information convex},''
\href{http://arxiv.org/abs/1810.01986}{{\ttfamily arXiv:1810.01986
  [cond-mat.str-el]}}.

\bibitem{lou2019ishibashi}
J.~Lou, C.~Shen, and L.-Y. Hung, ``Ishibashi states, topological orders with
  boundaries and topological entanglement entropy. part i,'' {\em Journal of
  High Energy Physics} {\bfseries 2019} no.~4, (2019) 17.

\bibitem{Hu:2019bec}
Y.~Hu and Y.~Wan, ``{Entanglement Entropy, Quantum Fluctuations, and Thermal
  Entropy in Topological Phases},''
  \href{http://dx.doi.org/10.1007/JHEP05(2019)110}{{\em JHEP} {\bfseries 05}
  (2019) 110},
\href{http://arxiv.org/abs/1901.09033}{{\ttfamily arXiv:1901.09033
  [cond-mat.str-el]}}.

\bibitem{Luo:2018yqb}
Z.-X. Luo, B.~G. Pankovich, Y.~Hu, and Y.-S. Wu, ``{Correspondence between bulk
  entanglement and boundary excitation spectra in two-dimensional gapped
  topological phases},''
  \href{http://dx.doi.org/10.1103/PhysRevB.99.205137}{{\em Phys. Rev.}
  {\bfseries B99} no.~20, (2019) 205137},
\href{http://arxiv.org/abs/1806.07794}{{\ttfamily arXiv:1806.07794
  [cond-mat.str-el]}}.

\bibitem{beigi_quantum_2011}
S.~Beigi, P.~W. Shor, and D.~Whalen, ``The {Quantum} {Double} {Model} with
  {Boundary}: {Condensations} and {Symmetries},''
  \href{http://dx.doi.org/10.1007/s00220-011-1294-x}{{\em Communications in
  Mathematical Physics} {\bfseries 306} no.~3, (Sept., 2011) 663--694}.
  \url{http://arxiv.org/abs/1006.5479}. 00092 arXiv: 1006.5479.

\bibitem{cong_topological_2016}
I.~Cong, M.~Cheng, and Z.~Wang, ``Topological {Quantum} {Computation} with
  {Gapped} {Boundaries},'' {\em arXiv:1609.02037 [cond-mat, physics:quant-ph]}
  (Sept., 2016) . \url{http://arxiv.org/abs/1609.02037}. arXiv: 1609.02037.

\bibitem{Hu:2017faw}
Y.~Hu, Z.-X. Luo, R.~Pankovich, Y.~Wan, and Y.-S. Wu, ``{Boundary Hamiltonian
  theory for gapped topological phases on an open surface},''
  \href{http://dx.doi.org/10.1007/JHEP01(2018)134}{{\em JHEP} {\bfseries 01}
  (2018) 134},
\href{http://arxiv.org/abs/1706.03329}{{\ttfamily arXiv:1706.03329
  [cond-mat.str-el]}}.

\bibitem{Hu:2017btw}
Y.~Hu, Y.~Wan, and Y.-S. Wu, ``{From effective Hamiltonian to anomaly inflow in
  topological orders with boundaries},''
  \href{http://dx.doi.org/10.1007/JHEP08(2018)092}{{\em JHEP} {\bfseries 08}
  (2018) 092},
\href{http://arxiv.org/abs/1706.09782}{{\ttfamily arXiv:1706.09782
  [cond-mat.str-el]}}.

\bibitem{Wang:2018qvd}
H.~Wang, Y.~Li, Y.~Hu, and Y.~Wan, ``{Gapped Boundary Theory of the Twisted
  Gauge Theory Model of Three-Dimensional Topological Orders},''
  \href{http://dx.doi.org/10.1007/JHEP10(2018)114}{{\em JHEP} {\bfseries 10}
  (2018) 114}, \href{http://arxiv.org/abs/1807.11083}{{\ttfamily
  arXiv:1807.11083 [cond-mat.str-el]}}.
[JHEP18,114(2020)].

\bibitem{petkova_generalised_2001}
V.~B. Petkova and J.-B. Zuber, ``Generalised twisted partition functions,''
  \href{http://dx.doi.org/10.1016/S0370-2693(01)00276-3}{{\em Physics Letters
  B} {\bfseries 504} no.~1-2, (Apr., 2001) 157--164}.
  \url{http://arxiv.org/abs/hep-th/0011021}. 00164 arXiv: hep-th/0011021.

\bibitem{shen2019defect}
C.~Shen and L.-Y. Hung, ``A defect verlinde formula,'' {\em arXiv preprint
  arXiv:1901.08285} (2019) .

\bibitem{dong_topological_2008}
S.~Dong, E.~Fradkin, R.~G. Leigh, and S.~Nowling, ``Topological entanglement
  entropy in {Chern}-{Simons} theories and quantum {Hall} fluids,''
  \href{http://dx.doi.org/10.1088/1126-6708/2008/05/016}{{\em Journal of High
  Energy Physics} {\bfseries 2008} no.~05, (2008) 016}.
  \url{http://stacks.iop.org/1126-6708/2008/i=05/a=016}. 00126.

\bibitem{fliss_interface_2017}
J.~R. Fliss, X.~Wen, O.~Parrikar, C.-T. Hsieh, B.~Han, T.~L. Hughes, and R.~G.
  Leigh, ``Interface {Contributions} to {Topological} {Entanglement} in
  {Abelian} {Chern}-{Simons} {Theory},''
  \href{http://dx.doi.org/10.1007/JHEP09(2017)056}{{\em Journal of High Energy
  Physics} {\bfseries 2017} no.~9, (Sept., 2017) }.
  \url{http://arxiv.org/abs/1705.09611}. 00012 arXiv: 1705.09611.

\bibitem{wen_edge_2016}
X.~Wen, S.~Matsuura, and S.~Ryu, ``Edge theory approach to topological
  entanglement entropy, mutual information and entanglement negativity in
  {Chern}-{Simons} theories,''
  \href{http://dx.doi.org/10.1103/PhysRevB.93.245140}{{\em Physical Review B}
  {\bfseries 93} no.~24, (June, 2016) }. \url{http://arxiv.org/abs/1603.08534}.
  00031 arXiv: 1603.08534.

\bibitem{Barkeshli:2013yta}
M.~Barkeshli, C.-M. Jian, and X.-L. Qi, ``{Theory of defects in Abelian
  topological states},''
  \href{http://dx.doi.org/10.1103/PhysRevB.88.235103}{{\em Phys. Rev.}
  {\bfseries B88} (2013) 235103},
\href{http://arxiv.org/abs/1305.7203}{{\ttfamily arXiv:1305.7203
  [cond-mat.str-el]}}.

\bibitem{lan_gapped_2015}
T.~Lan, J.~Wang, and X.-G. Wen, ``Gapped {Domain} {Walls}, {Gapped}
  {Boundaries} and {Topological} {Degeneracy},''
  \href{http://dx.doi.org/10.1103/PhysRevLett.114.076402}{{\em Physical Review
  Letters} {\bfseries 114} no.~7, (Feb., 2015) }.
  \url{http://arxiv.org/abs/1408.6514}. arXiv: 1408.6514.

\bibitem{blumenhagen_introduction_2009}
R.~Blumenhagen and E.~Plauschinn, {\em Introduction to conformal field theory:
  with applications to string theory}.
\newblock No.~779 in Lecture notes in physics. Springer, Dordrecht ; New York,
  2009.
\newblock OCLC: ocn310400947.

\bibitem{kapustin_topological_2011}
A.~Kapustin and N.~Saulina, ``Topological boundary conditions in abelian
  {Chern}-{Simons} theory,''
  \href{http://dx.doi.org/10.1016/j.nuclphysb.2010.12.017}{{\em Nuclear Physics
  B} {\bfseries 845} no.~3, (Apr., 2011) 393--435}.
  \url{http://arxiv.org/abs/1008.0654}. 00075 arXiv: 1008.0654.

\bibitem{levin_protected_2013}
M.~Levin, ``Protected edge modes without symmetry,''
  \href{http://dx.doi.org/10.1103/PhysRevX.3.021009}{{\em Physical Review X}
  {\bfseries 3} no.~2, (May, 2013) }. \url{http://arxiv.org/abs/1301.7355}.
  00109 arXiv: 1301.7355.

\bibitem{wang_boundary_2015}
J.~C. Wang and X.-G. Wen, ``Boundary degeneracy of topological order,''
  \href{http://dx.doi.org/10.1103/PhysRevB.91.125124}{{\em Physical Review B}
  {\bfseries 91} no.~12, (Mar., 2015) }.
  \url{https://link.aps.org/doi/10.1103/PhysRevB.91.125124}. 00041.

\bibitem{wen_topological_1995}
X.-G. Wen, ``Topological orders and {Edge} excitations in {FQH} states,''
  \href{http://dx.doi.org/10.1080/00018739500101566}{{\em Advances in Physics}
  {\bfseries 44} no.~5, (Oct., 1995) 405--473}.
  \url{http://arxiv.org/abs/cond-mat/9506066}. 00002 arXiv: cond-mat/9506066.

\bibitem{Bais:2002ny}
F.~A. Bais, B.~J. Schroers, and J.~K. Slingerland, ``{Hopf symmetry breaking
  and confinement in (2+1)-dimensional gauge theory},''
  \href{http://dx.doi.org/10.1088/1126-6708/2003/05/068}{{\em JHEP} {\bfseries
  05} (2003) 068},
\href{http://arxiv.org/abs/hep-th/0205114}{{\ttfamily arXiv:hep-th/0205114
  [hep-th]}}.

\bibitem{Bais:2002pb}
F.~A. Bais, B.~J. Schroers, and J.~K. Slingerland, ``{Broken quantum symmetry
  and confinement phases in planar physics},''
  \href{http://dx.doi.org/10.1103/PhysRevLett.89.181601}{{\em Phys. Rev. Lett.}
  {\bfseries 89} (2002) 181601},
\href{http://arxiv.org/abs/hep-th/0205117}{{\ttfamily arXiv:hep-th/0205117
  [hep-th]}}.

\bibitem{bais_condensate-induced_2009}
F.~A. Bais and J.~K. Slingerland, ``Condensate-induced transitions between
  topologically ordered phases,''
  \href{http://dx.doi.org/10.1103/PhysRevB.79.045316}{{\em Physical Review B}
  {\bfseries 79} no.~4, (Jan., 2009) 045316}.
  \url{https://link.aps.org/doi/10.1103/PhysRevB.79.045316}.

\bibitem{Bais:2008xf}
F.~A. Bais, J.~K. Slingerland, and S.~M. Haaker, ``{A Theory of topological
  edges and domain walls},''
  \href{http://dx.doi.org/10.1103/PhysRevLett.102.220403}{{\em Phys. Rev.
  Lett.} {\bfseries 102} (2009) 220403},
\href{http://arxiv.org/abs/0812.4596}{{\ttfamily arXiv:0812.4596
  [cond-mat.mes-hall]}}.

\bibitem{barkeshli_classification_2013}
M.~Barkeshli, C.-M. Jian, and X.-L. Qi, ``Classification of {Topological}
  {Defects} in {Abelian} {Topological} {States},''
  \href{http://dx.doi.org/10.1103/PhysRevB.88.241103}{{\em Physical Review B}
  {\bfseries 88} no.~24, (Dec., 2013) }. \url{http://arxiv.org/abs/1304.7579}.
  00058 arXiv: 1304.7579.

\bibitem{Kong:2013aya}
L.~Kong, ``{Anyon condensation and tensor categories},''
  \href{http://dx.doi.org/10.1016/j.nuclphysb.2014.07.003}{{\em Nucl. Phys.}
  {\bfseries B886} (2014) 436--482},
\href{http://arxiv.org/abs/1307.8244}{{\ttfamily arXiv:1307.8244
  [cond-mat.str-el]}}.

\bibitem{fuchs2013bicategories}
J.~Fuchs, C.~Schweigert, and A.~Valentino, ``Bicategories for boundary
  conditions and for surface defects in 3-d tft,'' {\em Communications in
  Mathematical Physics} {\bfseries 321} no.~2, (2013) 543--575.

\bibitem{hung_ground_2015}
L.-Y. Hung and Y.~Wan, ``Ground {State} {Degeneracy} of {Topological} {Phases}
  on {Open} {Surfaces},''
  \href{http://dx.doi.org/10.1103/PhysRevLett.114.076401}{{\em Physical Review
  Letters} {\bfseries 114} no.~7, (Feb., 2015) }.
  \url{http://arxiv.org/abs/1408.0014}. 00031 arXiv: 1408.0014.

\bibitem{hung_generalized_2015}
L.-Y. Hung and Y.~Wan, ``Generalized {ADE} {Classification} of {Gapped}
  {Domain} {Walls},'' \href{http://dx.doi.org/10.1007/JHEP07(2015)120}{{\em
  Journal of High Energy Physics} {\bfseries 2015} no.~7, (July, 2015) }.
  \url{http://arxiv.org/abs/1502.02026}. 00006 arXiv: 1502.02026.

\bibitem{cardy_boundary_1989}
J.~L. Cardy, ``Boundary conditions, fusion rules and the {Verlinde} formula,''
  \href{http://dx.doi.org/10.1016/0550-3213(89)90521-X}{{\em Nuclear Physics B}
  {\bfseries 324} no.~3, (Oct., 1989) 581--596}.
  \url{http://linkinghub.elsevier.com/retrieve/pii/055032138990521X}. 01113.

\bibitem{cardy_boundary_2004}
J.~Cardy, ``Boundary {Conformal} {Field} {Theory},'' {\em arXiv:hep-th/0411189}
  (Nov., 2004) . \url{http://arxiv.org/abs/hep-th/0411189}. 00339 arXiv:
  hep-th/0411189.

\bibitem{serre_cours_1994}
J.~P. Serre, {\em Cours d'arithmetique}.
\newblock 1994.

\end{thebibliography}\endgroup
    \bibliographystyle{utphys}
    \end{document}